%% file: 0.tex
\documentclass{llncs}

\usepackage{amssymb}
\usepackage{amsfonts}
\usepackage{amsmath}
\usepackage{graphicx}
\usepackage{bussproofs}
\usepackage{verbatim}
\usepackage{mathrsfs}
\usepackage{latexsym} 
\usepackage{color}
\usepackage{xspace}
\usepackage{floatflt}
\usepackage{float}
\usepackage[dvips,breaklinks=true, colorlinks=true, citecolor=blue, linkcolor=blue]{hyperref}
\usepackage{euscript}
\usepackage{xspace}
\usepackage{euscript}
\usepackage{textcomp}
\usepackage{enumerate}
\usepackage{ifthen}
\usepackage{tikz}

\usetikzlibrary{matrix}

\usetikzlibrary{automata,positioning}

\input my-macros.tex

\title{Guarded Variable Automata over Infinite Alphabets}

\begin{document}
\author{Walid Belkhir \inst{1} and Yannick Chevalier \inst{2} and Michael Rusinowitch \inst{1}}

\institute{%
  INRIA Nancy--Grand Est \& LORIA  \ 
  \email{walid.belkhir@inria.fr,rusi@loria.fr}
  \and
  Universit\'e Paul Sabatier \&  IRIT Toulouse \ 
  \email{ychevali@irit.fr}
}

\maketitle

\input abstract.tex
\input introduction.tex

\input preliminaries.tex

\input gfva.tex
\input fma.tex
\input properties.tex
\input emptiness.tex
\input simulation.tex
\input synthesis.tex

\input conclusion.tex

\bibliographystyle{abbrv}
\bibliography{biblio}

\newpage 
\input annex.tex

\end{document}

%% file: my-macros.tex
\def\Pos{\ensuremath{\text{\rm Pos}}}
\def\PosA{\ensuremath{\text{\rm Pos}}_{\texttt{A}}}
\def\PosE{\ensuremath{\text{\rm Pos}}_{\texttt{E}}}

\newcommand \AbelardM[1]{\big( #1 \big)_{\texttt{A}}}
\newcommand \EloiseM[2]{\big( #1, #2 \big)_{\texttt{E}}}

\newcommand  \blue[1]{\textcolor{blue}{#1}}

\newcommand \uberTo[1]{\stackrel{#1}{\rightarrow}}
\newcommand \UberTo[1]{\stackrel{#1}{\rTo}}
\newcommand \uberToR[1]{\stackrel{#1}{\Rightarrow}}
\newcommand \uberToST[1]{\stackrel{#1}{\Rightarrow}{^{\hspace{-2mm}\star}}}

\newcommand \bs[1] {\boldsymbol{#1}}

\newcommand \G  {\mathbb{G}}

\newcommand \ucomment[1]{}
\newcommand \unnecessary[1]{}
\newcommand \longversion[1]{}
\newcommand \myannex[1]{}

\newcommand \set[1]{\{{#1}\}}

\newcommand \citep[1]{\cite{#1}}

\newcommand \LongVersion[1]{}

\newcommand \tif {\text{ if }}

\newcommand \tand {\textrm{ and }}
\newcommand \tor {\text{ or }}
\newcommand \model[1]{\langle #1 \rangle}
\newcommand \ttt[1]{\texttt{#1}}

\newcommand \mcons{\ensuremath{\Sigma \cup \mathcal{X}}}

\newcommand {\tst}{ \textrm{ s.t. } }

\newcommand \Abelard {\texttt{Abelard}\xspace}
\newcommand \Eloise {\texttt{Eloise}\xspace}
\newcommand \N {\mathbb{N}}
\newcommand \ceil[1]{\lceil {#1} \rceil}

\newcommand {\GVAS} {$\textrm{GVA}^{\#}$} 

\newcommand {\vaut}[1] 
   {
  \ifthenelse{\equal{#1}{automaton}}{FVA\xspace}{}
  \ifthenelse{\equal{#1}{automata}}{FVAs}{}
  }

\newcommand {\Eu}[1]{\EuScript{#1}}
\newcommand {\eps} {\varepsilon}

\newcommand{\mycal}[1]{\mathcal{#1}}
\newcommand{\rTo}{\longrightarrow}
\newcommand \pfun {\rightharpoonup} 

\newcommand{\synch}{\Join}

\newcommand{\bigland}{\bigwedge}

\newcommand{\gvert}{\;\;|\;\;}

\newcounter{casecounter}
\def\thecasecounter{(\roman{casecounter})}

%% file: abstract.tex
\begin{abstract}
  We define \emph{guarded variable automata} (GVAs), a simple extension of finite automata
  over infinite alphabets. 
  In this model  the  transitions  are labeled by letters or variables ranging over an infinite alphabet and guarded 
  by conjunction of equalities and disequalities. 
GVAs are well-suited for modeling  component-based  applications such as web services.  
  They are closed  under  intersection,  union,  concatenation and  Kleene operator, 
  and their  nonemptiness problem is PSPACE-complete. 
  We show that the simulation preorder  of  GVAs  is decidable.  
Our proof relies  on the characterization of    the simulation by means of games and strategies.
This result can be  applied  to service composition synthesis.  
\end{abstract}


%% file: introduction.tex
\section{Introduction}
The simple and  powerful formalism of finite automata is  widely used for system specification and verification. 
Considerable  efforts have been  devoted to  extend  
finite automata  to  infinite alphabets: finite memory automata \cite{tcs/KaminskiF94}, 
data automata \cite{BojanczykLICS06},  
variable automata \cite{GKS10}, usage automata \cite{DFM:nominal:CIAA12}, 
fresh-variable automata \cite{BCR:FVA:13}, only 
to cite  a few (see \cite{amaldev2011}  for a survey). When developing formalisms over infinite alphabets, the main challenge 
is  to preserve as much as possible useful properties such as  compositionality (i.e. closure under  basic operations)
and  decidability of  basic problems  such as   nonemptiness, membership, universality,  language containment, simulation, etc   

The  language containment problem  is a particularly important one in applications 
like formal verification. 
For instance,   whether  an  implementation is conform 
to a specification amounts to decide the containment  $L(A) \subseteq L(B)$, where 
$A$ (resp. $B$) is an automaton formalizing the behavior  of the implementation (resp. specification),
and $L(A)$ is the language of words recognized by $A$.  

The  containment  problem for finite automata (FAs) can be solved 
by using determinization, 
in a   complete but inefficient way. Moreover, 
for several  classes of automata  over infinite alphabets, the containment problem turned out to be 
undecidable. This is the case for finite memory automata~\cite{NevenSV04} and variable automata \cite{GKS10}.  
As a practical alternative  approach, 
a simulation preorder can be employed to underapproximate the  containment relation (e.g.~\cite{Dill91}).
Indeed, simulation-based techniques are sometimes more  efficient. For instance 
a simulation between two  finite automata can be computed in polynomial time. 
To our knowledge,  simulation  has not been studied for the classes of automata 
over infinite alphabets from \cite{tcs/KaminskiF94} and \cite{GKS10}. 

Our work is also motivated by the  composition 
synthesis problem for web services in which the agents (i.e. client and the available services)
exchange data ranging over an infinite domain.
One of the most successful approaches to composition amounts to
abstract services as finite-state automata (FA) and apply available
tools from automata theory to synthesize a new service satisfying the
given client requests from an existing community of
services (e.g. \cite{BerardiCGP08,MW08}). In this setting 
synthesizing a new service amounts to compute a simulation relation 
of the client  by the community of the available services, e.g. \cite{BerardiCGP08}. 
However  it is not obvious whether   the automata-based  approach to service composition 
can still be applied with infinite alphabets since simulation often gets undecidable 
in extended models like Colombo (e.g.~\cite{AkrounBNT12}).
Following the approach initiated in \cite{BCR:FVA:13}
our objective  is  to  define 
expressive classes  of automata on infinite alphabets
which are   well-adapted to the specification and composition of services  
and   enjoy   nice closure  properties and decidable simulation.

\noindent \textbf{\emph{Contributions.}}  In this paper we define
\emph{guarded variable automata}, or GVAs, a natural extension of finite
automata over infinite alphabets.  In this model the transitions are
labeled by letters or variables ranging over an infinite alphabets and
guarded by conjunction of equalities and disequalities.
Besides, some variables are refreshed in some states, that is, these
variables can be released so that new letters can be bound to them.
The potential applicability of our model in verification (e.g. model
checking~\cite{Berard:2010:SSV:1965314}) and service composition
\cite{AkrounBNT12} follows from the fact that GVAs are closed
under 
intersection, union, concatenation and Kleene operator.  The
nonemptiness problem is shown to be PSPACE-complete for GVAs, and the membership is
NP-Complete.  However, their universality and containment problems are
undecidable.  We introduce a simulation preorder for GVAs and show its 
decidability.  The proof relies on a game-theoretic characterization
of simulation.

\noindent \textbf{\emph{Related work.}} 
GVAs are closely  related to the classes of automata in~\cite{tcs/KaminskiF94,GKS10,BCR:FVA:13}.
but these classes are strictly included in GVAs. We 
show below that GVAs have the same expressivity as 
\emph{finite-memory automata with non-deterministic reassignment} (NFMAs)  \cite{ijfcs:KaminskiZ10}.
Here we give the complexity status of nonemptyness for GVAs. 
This problem was not considered for NFMA in \cite{ijfcs:KaminskiZ10}.
We also give a procedure to decide the simulation preorder for GVAs. Simulation has not been studied  in \cite{ijfcs:KaminskiZ10,tcs/KaminskiF94,GKS10} and considered  only for the less expressive FVAs in \cite{BCR:FVA:13}. 

\noindent \textbf{\emph{Paper organization.}}
Sec. \ref{prelim:sec} recalls standard notions. 
Sec. \ref{gaut:sec} introduces the new  class of  guarded variable automata.
Subsec. \ref{subsec:gva:nfma:equiv} studies the expressiveness of GVAs  with respect to NFMAs.
Sec. \ref{prop:sec}  studies  closure properties and the complexity of Nonemptiness for GVAs. 
Sec. \ref{simulation:sec} introduces  the simulation preorder of  GVAs.
Sec. \ref{decidability:sec} shows its decidability.
Sec. \ref{service:synthesis} applies these results to service composition.
Future work directions  are given in Sec. \ref{conclusion:sec}.     
Missing proofs are provided in  external appendices.

%% file: preliminaries.tex
\section{Preliminaries}
\label{prelim:sec}

Let $\mathcal{X}$ be a finite set of variables, $\Sigma$ an infinite
alphabet of letters.  A substitution is an idempotent mapping
$\{x_1\mapsto \alpha_1,\ldots,x_n\mapsto \alpha_n\}\cup
\bigcup_{a\in\Sigma}\set{a \mapsto a }$ with variables $x_1, \ldots,
x_n$ in $\mathcal{X}$ and $\alpha_1, \ldots, \alpha_n$ in $\mycal{X}
\cup \Sigma $. We call $\set{x_1,\ldots,x_n}$ its \emph{proper
  domain}, and denote it by $dom(\sigma)$. We denote by $Dom(\sigma)$
the set $dom(\sigma) \cup \Sigma$.  We denote by $codom(\sigma)$ the
set 
 $\set{a\in \Sigma
  \gvert \exists x \in dom(\sigma) \tst \sigma(x)=a}$.  If all the
$\alpha_i,i=1\ldots n$ are letters then we say that $\sigma$ is
ground.  The empty substitution (\textit{i.e.}, with an empty proper
domain) is denoted by $\emptyset$.  The set of substitutions from
$\mycal{X}\cup\Sigma$ to a set $A$ is denoted by
$\zeta_{\mycal{X},A}$, or by $\zeta_{\mycal{X}}$, or simply by $\zeta$
if there is no ambiguity.  If $\sigma_1$ and $\sigma_2$ are
substitutions that coincide on the domain $dom(\sigma_1)\cap
dom(\sigma_2)$, then $\sigma_1 \cup \sigma_2$ denotes their union in
the usual sense.  If $dom(\sigma_1)\cap dom(\sigma_2)=\emptyset$ then we 
  denote by $\sigma_1 \uplus \sigma_2$ their \emph{disjoint} union.
We define the function $\mycal{V}:\Sigma\cup
\mycal{X} \rTo \mycal{P}(\mycal{X})$ by
$\mycal{V}(\alpha)=\set{\alpha}$ if $\alpha \in \mycal{X}$, and
$\mycal{V}(\alpha)=\emptyset$, otherwise.  For a function $F: A
\uberTo{} B$, and $A'\subseteq A$, the restriction of $F$ on $A'$ is
denoted by $F_{|A'}$. If $k\in \mathbb{N}$ then we let $[k]=\set{1,\ldots,k}$.

A two-players game is a tuple $\langle \Pos_{E},\Pos_{A},M, p^{\star}
\rangle$, where $\Pos_{E},\Pos_{A}$ are  disjoint sets of
positions: \Eloise's positions and \Abelard's positions. $M\subseteq
(\Pos_{E} \cup \Pos_{A}) \times (\Pos_{E} \cup \Pos_{A})$ is a set of
\emph{moves}, and $p^{\star}$ is the starting position.  A strategy
for the player \Eloise is a function $\rho: \Pos_E \uberTo{} \Pos_E \cup
Pos_A$, such that $(\wp,\rho(\wp)) \in M$ for all $\wp \in \Pos_E$. A
(possibly infinite) play $\pi=\langle \wp_1,\wp_2,\ldots \rangle$
\emph{follows} a strategy $\rho$ for player \Eloise iff
$\wp_{i+1}=\rho(\wp_i)$ for all $i \in \mathbb{N}$ such that $\wp_i
\in \Pos_{E}$. Let $\Eu{W}$ be a (possibly infinite) set of plays. 
A strategy $\rho$ is \emph{winning} for \Eloise from a
set $S \subseteq \Pos_E\cup \Pos_A$ according to   $\Eu{W}$
iff every play starting from a
position in $S$ and following $\rho$ belongs to $\Eu{W}$.

%% file: gfva.tex
\section{Guarded variable automata}
\label{gaut:sec}
In this section we define formally  the class of GVAs.  
It is an extension of FVAs  \cite{BCR:FVA:13} 
with logical constraints, called guards. 

 Let us first explain  the main ideas behind GVAs. 
 The transitions of a GVA  are labeled with letters or  variables ranging over an infinite set of letters. 
 These transitions can  also be  labeled with guards consisting of equalities and disequalities.  
 Its guard must be true for the transition to be fired. We emphasize  that 
 while reading a guarded transition some  variables of the guard might be free and we need to \emph{guess}
 their value.   
 Finally, some variables are refreshed in some states, that is, variables  
 can be \emph{freed} in these states so that  new letters can be assigned to  them.
 Firstly, we introduce the syntax and semantics of guards.

\begin{definition}
  The set $\G$ of guards over $\Sigma \cup \mycal{X}$ is inductively defined as follows:
  $G  \; :=\;   \ttt{\emph{true}} \gvert \alpha=\beta \gvert \alpha \neq \beta \gvert  G \land G$,
  where  $\alpha,\beta \in \Sigma \cup \mycal{X}$. 
  We write $\sigma \models g$ if a substitution $\sigma$ satisfies a guard $g$. 
\end{definition}

We notice that adding the disjunction operator to  the guards would   not 
increase the expressivity of our  model. 
A guard is atomic iff it is either 
\ttt{true}, an equality, or an inequality. 
Let $g_i,i=1,\ldots, n$,  be atomic guards. Then define the \emph{free} 
 variables of a  guard by 
 $\mycal{V}(\bigland_{i=1,n} g_i)=\bigcup_{i=1,n} \mycal{V}(g_i)$ and 
$\mycal{V}(\alpha \sim \beta )= \mycal{V}(\alpha) \cup  \mycal{V}(\beta))$, where $\sim \in \set{=,\neq}$ 
and $\alpha,\beta \in \Sigma \cup \mycal{X}$. 
The application of a substitution $\gamma$ to a guard $g$, denoted by $\gamma(g)$,
is defined in the usual way. 
The formal definition of GVAs follows.
\begin{definition}
  A \emph{GVA} is a tuple
  $\mycal{A}=\model{\Sigma,\mathcal{X},Q,Q_0,\delta,F,\kappa}$ where $\Sigma$
  is an infinite set of letters, $\mathcal{X}$ is a finite set of
  variables, $Q$ is a finite set of states, $Q_0\subseteq Q$ is a set
  of initial states, 
  $\delta:Q \times (\Sigma_{\mycal{A}} \cup \mycal{X} \cup \set{\eps}) \times \G\to 2^{Q}$ is a
  transition function where $\Sigma_{\mycal{A}}$ is a finite subset of $\Sigma$,
  $F\subseteq Q$ is a set of accepting states,
  and $\kappa: \mycal{X} \rightarrow 2^Q$ is  called   the refreshing  function. 
\end{definition}

The run of a GVA  is defined over \emph{configurations}. A configuration is 
a pair $(\gamma,q)$ where $\gamma$ is a substitution such that for all variables 
$x$ in $dom(\gamma)$, $\gamma(x)$ is  the current value of $x$, 
and $q$ is a state of the GVA.   
Intuitively,  when a GVA $\mycal{A}$ is in state $q$, and  $(\gamma,q)$ is the current configuration, 
  and there is  a transition $q \uberTo{\alpha,g} q'$  in $\mycal{A}$ then: 

\begin{itemize}
\item if $\alpha$ is a free variable (i.e. $\alpha \in \mathcal{X} \setminus dom(\gamma)$) then $\alpha$ stores the input  letter  and  some values for all the other  free 
variables of $\gamma(g)$ are  \emph{guessed} such that $\gamma(g)$ holds, 
        and $\mycal{A}$ enters   state $q' \in \delta(q,\alpha,g)$, 
\item if $\alpha$ is a bound variable or a letter (i.e. $\alpha \in Dom(\gamma)$) and  
$\gamma(\alpha)$  is equal to  the input  letter  $l$ then  some values for  all the 
         free variables of $\gamma(g)$ are \emph{guessed} such  that  $\gamma(g)$ holds, and 
         $\mycal{A}$ enters   state $q'\in \delta(q,\alpha,g)$.
\end{itemize}
In both cases  when $\mycal{A}$ enters state $q'$ all the variables which are 
refreshed in $q'$ are freed. Thus the purpose of guards is to compare letters
and to guess new letters that might be read afterward. 

For a GVA $\mycal{A}$, we shall denote by $\Sigma_{\mycal{A}}$ 
the finite set of letters 
that appear in the transition function of
$\mycal{A}$. We shall denote by   $\kappa^{-1}: Q \uberTo{} 2^{\mycal{X}}$ the function 
 that associates to each state of the GVA the set of variables being refreshed in this state. 
That is, $\kappa^{-1}(q)=\set{x \in \mycal{X} \gvert q \in \kappa(x)  }$.

\noindent The formal definitions of configuration, run and recognized language follow.
\begin{definition}
  \label{run:lang:def}
  Let   $\mycal{A}=\model{\Sigma,\mathcal{X},Q,Q_0,\delta,F,\kappa}$ be a GVA.  
  A \emph{configuration}  is a pair  $(\gamma,q)$ where   %
$\gamma$  is a substitution 
 and $q \in Q$. 
  We define a transition relation  over the  configurations 
  as follows:  $(\gamma_1,q_1)\uberToR{a} (\gamma_2,q_2)$, where $a \in \Sigma \cup \set{\eps}$,
  iff there exists a substitution $\sigma$   such that  $dom(\sigma)\cap dom(\gamma_1)=\emptyset$
  and either:  
\begin{enumerate}[i)]
\item  $a \in \Sigma$ and in this case there exists a  label $\alpha \in \Sigma\cup \mycal{X}$ 
      such that  $q_2\in \delta(q_1,\alpha,g)$,
     $(\gamma_1 \uplus \sigma)(\alpha)=a$,
     $(\gamma_1\uplus \sigma)\models g$ and 
    $\gamma_2=(\gamma_1\uplus \sigma)_{| D}$, with $D= Dom(\gamma_1 \uplus \sigma )\setminus  \kappa^{-1}(q_2)$.  Or,
\item $a = \eps$ and in this case 
     $(\gamma_1\uplus \sigma)\models g$ and 
    $\gamma_2=(\gamma_1\uplus \sigma)_{| D}$, with $D= Dom(\gamma_1 \uplus \sigma )\setminus  \kappa^{-1}(q_2)$.
\end{enumerate}
 We denote by $\uberToST{}$ the reflexive and transitive closure of $\uberToR{}$. For two configurations 
  $\bs{c}, \bs{c}'$ and a letter $w\in \Sigma$, we write 
  $\bs{c} \uberTo{w} \bs{c}'$ iff  there exists two configurations $\bs{c}_1$ and $\bs{c}_2$ such that 
    $\bs{c} \uberToST{\eps} \bs{c}_1 \uberToR{w} \bs{c}_2 \uberToST{\eps} \bs{c}'$.
  A finite word $w=w_1w_2\ldots w_n \in \Sigma^*$ is \emph{recognized} by 
  $\mycal{A}$ iff  there exists a run
  $(\gamma_{0},q_0) \uberTo{w_1}(\gamma_1,q_1)\uberTo{w_2} \ldots \uberTo{w_n}(\gamma_n,q_n)$, such that 
  $q_0\in Q_0$ and $q_n \in F$.
  The set of words recognized by $\mycal{A}$ is denoted by
  $L(\mycal{A})$.
\end{definition}

\hspace{-0.55cm}
\begin{minipage}{.60\textwidth}
\begin{example}
\label{GFVA:ex}
 Let $\mycal{A}_1$ 
 and $\mycal{A}_2$ be the GVAs depicted below  where $x,y$ are variables and 
 $\kappa_i$ the refreshing function of $\mycal{A}_i$, $i=1,2$, is defined by 
$\kappa_1(y)=\set{p_0}$ and $\kappa_2(x)=\kappa_2(y)=\set{q_0}$.
We notice that while making the first loop over
$p_0$, 
  the variable $x$ of the guard $(y\neq x)$ is free and its value is guessed.
 Then the variable $y$ is refreshed in $p_0$, and at each loop 
 the input   letter should be different than the value of the variable 
 $x$ already guessed.  
\end{example}
\end{minipage}
\begin{minipage}{.40\textwidth}
\hspace{.7cm}
\begin{tikzpicture}[shorten >=1pt,node distance=2cm, bend angle=60,
            on grid,auto, initial text=, >=stealth] 
  \begin{scope}[yshift=-2cm]
          \node[state,initial] (r_0)   {$p_0$}; 
          \node[state,accepting] (r_1) [right =of r_0] {$p_1$}; 
          \path[->] 
          (r_0)  edge   node [above] {$x$} (r_1)
          (r_0) edge [loop above]  node [above] {$y, y\neq x$} ()
          -- (-0.9,0.6) node {$\mycal{A}_1$};
\end{scope}
          \begin{scope}[xshift=0cm, yshift=-3.6cm,shorten >=1pt,node distance=2cm, bend angle=50,
              on grid,auto, initial text=, >=stealth] 
            \node[state,initial,accepting] (r_0)   {$q_0$}; 
            \node[state] (r_1) [right =of r_0] {$q_1$}; 
            \path[->] 
            (r_0) [bend left] edge   node [above] {$x$} (r_1)
            (r_1) [bend left] edge  node  [below] {$y, y \neq x$} (r_0)
            -- (-0.9,0.6) node {$\mycal{A}_2$}; 
          \end{scope}
      \end{tikzpicture}
\end{minipage}
 Hence, the language  $L(\mycal{A}_1)$ consists of
all the words in $\Sigma^{\star}$ in which the last letter 
  is different than all the other letters. This language can be recognized
by a variable automaton \cite{GKS10} and by a NFMA \cite{ijfcs:KaminskiZ10} 
 but not by a FMA
\cite{tcs/KaminskiF94}.
On the other hand, the language $L(\mycal{A}_2)=\{w_1w'_1\cdots w_{n}w'_n \gvert 
w_i, w'_i \in \Sigma, \, n\ge 1, \tand w_{i} \neq w'_{i}, \, 
\forall i \in [n]\}$ can be recognized by a FMA but not by a variable
automaton.


%% file: fma.tex
\section{Comparison between GVAs and NFMA}
\label{subsec:gva:nfma:equiv}


In this section we show that  GVAs and  NFMAs recognize the same languages.
 We recall that a NFMA \cite{ijfcs:KaminskiZ10}
is a $8$-tuple $\mycal{F}=\model{\Sigma,k,Q,q_0,\bs{u},\rho,\delta,F}$ where
$k\in \mathbb{N}^{+}$ is the number of registers,
$Q$ is a finite set of states, 
$q_0 \in Q$ is the initial state, 
$\bs{u}: [k] \pfun \Sigma$ is a partial function called the \emph{initial assignment} of the $k$ registers,
$\rho: \set{(p,q) : (p,\eps,q) \in \delta} \pfun [k]$ 
is a  function  called the \emph{non-deterministic reassignment}, 
$\delta: Q\times ([k]\cup \set{\eps}) \times Q$ is  the \emph{transition relation},
  and $F \subseteq Q$ is the set of final states. 
Intuitively, if $\mycal{F}$ is in state $p$, and there is an $\eps$-transition
from $p$ to $q$ and $\rho(p,q)=l$, then $\mycal{F}$ can  non-deterministically replace the 
content of the $l^{\rm{th}}$ register with an element of $\Sigma$ not occurring in
any other register and enter state $q$. 
However, if $\mycal{F}$ is in state $p$, and the input symbol is equal to the content
of the $l^{\rm{th}}$  and $(p,l,q) \in \delta$ then $\mycal{F}$ may enter state $q$ and 
pass to the next input symbol.  
An $\eps$-transition $(p,\eps,q)\in \delta$ with $\rho(p,q)=l$,  for a register $l \in [k]$,    
is  denoted by $(p,\eps \slash l, q)$. 

\begin{figure}[H]
\vspace{-9mm}
\centering
\begin{tikzpicture}[shorten >=1pt,node distance=1.7cm, bend angle=60,
            on grid,auto, initial text=, >=stealth] 
  \begin{scope}
          \node[state,initial,initial distance=1.25cm] (p_0)   {$p$}; 
          \node[state] (p_1) [below left =of p_0] {$p'$}; 
          \node[state] (p_2) [below right  =of p_0] {$p''$};

          \path[->] 
          (p_0)  edge   node [left] {$m$} (p_1)
          (p_0)  edge   node [right] {$\eps\slash l$} (p_2)

          -- (0.16,-2.2) node {NFMA $\mycal{A}$}
          -- (2.4,-0.4) node {\begin{Large}$\leadsto$\end{Large}};
\end{scope}
          \begin{scope}[xshift=5.3cm, yshift=0cm,shorten >=1pt, bend angle=50,
              on grid,auto, initial text=, >=stealth] 

          \node[state,initial,initial distance=1.25cm] (q_0)   {$p$}; 
          \node[state] (q_1) [below left =of q_0] {$p'$}; 
          \node[state] (q_2)  [right  =of q_0] {$\tilde{p}$}; 
          \node[state] (q_3) [below right  =of q_2] {$p''$}; 

          \path[->] 
          (q_0)  edge   node [left] {$x_m$} (q_1)
          (q_0)  edge   node [above] {$\eps$} (q_2)
          (q_2)  edge   node [right] {$\eps,\bigland_{i\in[k]\setminus\set{l}}(x_l \neq x_i) $} (q_3)

            -- (2.1,0.76) node {$\kappa(x_l)=\set{\tilde{q}}$}
            -- (0.6,-2.2)  node {GVA $\mycal{A}'$}; 
          \end{scope}
      \end{tikzpicture}
\caption{A translation of NFMA to GVA. 
  The registers of the NFMA $\mycal{A}$ are $\set{1,\ldots,k}$, 
   they correspond to   the variables $\set{x_1,\ldots,x_k}$ of the GVA $\mycal{A}'$.
   The variable $x_l$ is refreshed     in the state $\tilde{q}$ of  $\mycal{A}'$.}
\label{translate:NFMA:GVA:fig}
\end{figure}
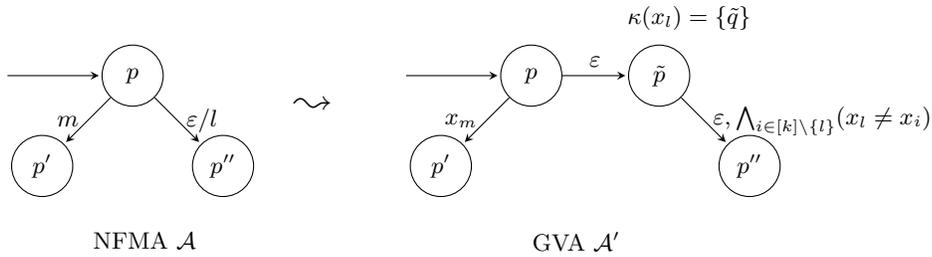

On the one hand, we can show that there is a polynomial  time translation of 
 a  NFMA  (with $k$ registers) into a GVA (with $k$ variables) of linear size and 
recognizing  the same language. 
More precisely, as shown in   Figure \ref{translate:NFMA:GVA:fig}: 

\begin{itemize}
\item a transition $(p,m,p')$ of the NFMA is translated as such, i.e. to $(p,x_m,p')$
\item a transition $(p,\eps\slash l,p'')$ of the NFMA is translated 
      to  two transitions $(p,\eps,\tilde{p})$ and $(\tilde{p},(x_l,g),p'')$ where 
$g=\bigland_{i \in [k]\setminus \set{l}} (\eps \neq x_i) $ and $x_l$ is refreshed in state $\tilde{p}$.
\end{itemize}

On the other hand, we show next that 
 a GVA can be translated into a NFMA 
 recognizing the same language.   
The idea is that the  $\eps$-transitions of the NFMA are used 
to encode the refreshing of the variables of the GVA.
For this purpose we introduce an intermediary class of GVAs, called \GVAS,  
in which 
the variables should have  distinct values. 
Then we translate \GVAS into NFMA.

\begin{definition}
Let \GVAS be the subclass of GVAs such that every  $\mycal{A}$ in \GVAS, 
verifies  $i)$  $\mycal{A}$ has no constants, i.e. $\Sigma_{\mycal{A}}=\emptyset$,  and
$ii)$ for every accessible configuration $(\sigma,q)$ of $\mycal{A}$ and  
  for all $x,y \in dom(\sigma)$,    $\sigma(x) \neq \sigma(y)$.
\end{definition}

We show next that GVAs and \GVAS{s} recognize the same language, 
more precisely we have:
\begin{lemma}
\label{GVA:GVAS:lemma}
For every GVA $\mycal{A}$ with $k$ variables and $n$ states 
there is a \GVAS  with $k+m$ variables and $O( n \cdot (k+m)!)$ states 
recognizing the same languages, where $m=|\Sigma_{\mycal{A}}|$.
\end{lemma}

Every \GVAS can be turned into a NFMA recognizing the same languages
by encoding the refreshing of the variables of the \GVAS with 
$\eps$-transitions. Hence,   

\begin{corollary}
For every \GVAS with $k$ variables  and $n$ states 
there exists a NFMA with $n \cdot k!$ states recognizing the same languages.
\end{corollary}

\input succinctness.tex


%% file: succinctness.tex
\subsection{Succinctness of GVAs w.r.t. NFMAs}

Though  GVAs and NFMAs  recognize the same languages, the latter 
are less adapted for service specification since their global
requirement that distinct variables must have distinct instances 
is an obstacle to compositionality. 
Assume for instance that we want to compose a payment service $P(c)$
with a reservation service $R(c')$ where $c$ (resp. $c'$) is the 
payer (resp. traveller) name taking value in some infinite alphabet.  If the services are specified as NFMAs 
a mediator (i.e. a service that delegates any client action to an appropriate service) would have to interact 
with two copies of the payment service, one in which the payer name is the same as the
traveler, and one in which they differ.
The choice by the mediator 
of one copy anticipates on the equalities that will be imposed by
future messages (The mediator is non-deterministic in Milner's weak determinacy sense~\cite[\S 11, Def 3]{milnerdeterminacy}). 
Since GVAs do not impose this early (non-deterministic) choice, 
they are more suited to  service specification and composition.

\subsection{Weak Determinacy}
\label{subsec:weak:determinacy}

In the rest of this section, a GVA or NFMA configuration is a couple
$s,m$ where $s$ is an automaton state and $m$ is an assignment to
variables or registers. Given an automaton $\mathcal{A}$, we denote
$s,m\to^\varepsilon_\mathcal{A} s',m'$ if there is an
$\varepsilon$-transition from the configuration $s,m$ to the
configuration $s',m'$, and $s,m\to^x_\mathcal{A} s',m'$ if there
exists a transition (whose guard is satisfied) from $s,m\to^x s',m'$
labeled with the variable, register, or letter $x$. We define the
relation $\Rightarrow^x_\mathcal{A}$ (reading of a register or letter
$x$) as $(\to^\varepsilon_\mathcal{A} * \circ \to^x_\mathcal{A})$. A
\emph{run} of $\mathcal{A}$ is a sequence of configurations
$(s_i,m_i)_{0\le i\le n}$ such that for all $0\le i < n$ we have
$s_i,m_i \Rightarrow^{x_{i+1}}_\mathcal{A} s_{i+1},m_{i+1}$ such that
$s_0$ is an initial state and $m_0$ is a possible initial
assignment. It is \emph{accepting} if $s_n$ is a final state. 
Finally we say that a run $s_0,m_0
\Rightarrow^{x_1}\ldots\Rightarrow^{x_n}$ \emph{reads} a word
$\omega=\omega_1 \cdot\ldots\cdot \omega_n$ if for all $1\le i\le n$
we have $\omega_i=m_i(x_i)$.
In the
rest of this section we omit the subscript in
$\Rightarrow_\mathcal{A}$ when the automaton $\mathcal{A}$ is clear
from the context. First let us define \emph{active variables} as the
subset of variables whose values has not changed since the last time
they were read.

\begin{definition}{\label{def:active:variables}(Active variables)}
  In a run $s_0,m_0 \Rightarrow^{x_1}\ldots\Rightarrow^{x_n}s_n,m_n$, the set
  $A_i$ of variables \emph{active in state $i$} is:
  \begin{itemize}
  \item $\emptyset$ if $i=0$
  \item $(A_{i-1}\setminus R_i)\cup\lbrace x_i\rbrace$ where $R_i$ is
    the set of variables or registers $x$ such that $m_{i-1}(x)\neq m_{i}(x)$. 
  \end{itemize}
\end{definition}

Since the automata we consider are fundamentally non-deterministic
(\textit{e.g.}, in guessing a new value for a register) we introduce a
notion of \emph{weak determinacy} adapted
from~\cite{milnerdeterminacy}. 

\begin{definition}{\label{milnerdeterminacy}(Weakly deterministic automata)}
  An automaton is \emph{weakly deterministic} if for every word
  $\omega$, if $s_0,m_0
  \Rightarrow^{x_1}\ldots\Rightarrow^{x_n}s_n,m_n$ and $s'_0,m'_0
  \Rightarrow^{x_1}\ldots\Rightarrow^{x_n}s'_n,m'_n$ are two runs
  reading $\omega$ then:
  \begin{itemize}
  \item We have $s_n=s'_n$
  \item A variable is active in $s_n,m_n$ iff it is active in $s_n,m'_n$;
  \item For every active variable $x$ in $s_n,m_n$ we have
    $m_n(x)=m'_n(x)$.
  \end{itemize}
\end{definition}

We remark that all the examples given in~\cite{ijfcs:KaminskiZ10} are weakly
deterministic NFMA.

\subsection{Succinctness}
\label{subsec:succinctness}

We have seen that it is possible to translate in polynomial time an
NFMA into a GVA that recognizes the same language. Conversely, we
prove in this section that there exists a sequence of languages
$(\mathcal{L}_n)_{n\ge 0}$ that each can be recognized by a weakly
deterministic GVA of size $O(n^2)$ and a weakly deterministic NFMA of
size $\Omega(n^n)$. This justifies our assertion that GVA can be
exponentially more succinct than NFMA in the class of weakly
deterministic automata. 

\paragraph{Definition of the $\lbrace \mathcal{L}_n\rbrace_{n\ge 0}.$}
We consider a sequence of languages of words of finite length which
are essentially palindroms, with an additional structure added to cope
with the non-determinism of NFMA. We denote $\tilde\omega$ the mirror
of a word $\omega$, and $\vert\omega\vert_a$ the number of $a$
occurring in $\omega$, for $a\in\Sigma$. We define, for $n\ge 0$:
$$
\mathcal{L}_n= \left\lbrace \underbrace{a_1\cdot\ldots\cdot
    a_n}_{\omega_1} \cdot \underbrace{b_1 \cdot\ldots\cdot
    b_n}_{\omega_2} \cdot \# \cdot\underbrace{ b_n \cdot \ldots \cdot
    b_1}_{\tilde\omega_2} \cdot \underbrace{a_n \cdot \ldots\cdot
    a_1}_{\tilde\omega_1} \vert \forall a, \vert \omega_2\vert_a > 0 \Rightarrow
  \vert \omega_1\vert_a > 0 \right\rbrace
$$
In order to simplify the statements, we say that in a run reading a
word of $\mathcal{L}_n$ and ending in a final state:
\begin{itemize}
\item The \emph{initialization phase} is the sub-run in which
  $a_1\cdot\ldots\cdot a_n$ is read (including the possible
  $\varepsilon$-transitions);
\item The \emph{pivot} is the state reached after reading $b_n$ for
  the first time;
\end{itemize}

\begin{proposition}
  Each $\mathcal{L}_n$ is recognized by a GVA of $4n+2$ states and
  $1+3n+n^2$ transitions.
\end{proposition}

\begin{proof}
  The constructed automaton has $2n$ variables, denoted
  $x_{a_1},\ldots,x_{a_n},x_{b_1},\ldots,x_{b_n}$ and is linear. The
  first $n$ states read and instantiate the variables
  $x_{a_1},\ldots,x_{a_n}$, the $n$ following states read and
  instantiate the variables $x_{b_1},\ldots,x_{b_n}$, with $n$
  available transitions from each state to its successor. For $1\le j
  \le n$ there exists a transition reading $b_i$ guarded with
  $x_{b_i}=x_{a_j}$. The remaining $2n+1$ states and transitions check
  that the word belongs to the language by reading in sequence (and
  without refreshment)
  $\#,x_{b_n},\ldots,x_{b_1},x_{a_n},\ldots,x_{a_1}$. The last state
  is the final state. The last state is a final state.
\end{proof}

Let us now prove that any NFMA recognizing $\mathcal{L}_n$ must have
at least $n^n$ states. Given a word $\omega$ of length $n$ in which
each letter occurs at most once (\textit{i.e.}, for all $a\in\Sigma$
we have $0\le \vert\omega\vert_a\le 1$), we let $L_\omega\subseteq
\mathcal{L}_n$ be the subset of $\mathcal{L}_n$ of words beginning
with $\omega$. The following lemmatas hold for all NFMA. The notation
$s,m$ stands for being in a state $s$ with the assignment to registers
$m$, and $x$ is either an $\varepsilon$-transition (possibly with
guessing) or the reading of a register or a constant. The notation
$\delta_{a,b}$ denotes the replacement of $a$ by $b$ in a
configuration or in a transition (if it isn't an $\varepsilon$ one).

\begin{lemma}
  If $s,m \longrightarrow^x s',m'$, $a\in \text{codom}(m)$ and $b
  \notin \text{codom}(m) \cup\text{codom}(m') \cup \lbrace x\rbrace$
  then $s,m\delta_{a,b} \longrightarrow^x s',m'\delta_{a,b}$.
\end{lemma}

\begin{proof}
  The proof proceeds by case analysis:
  \begin{itemize}
  \item If $x$ is an $\varepsilon$-transition:
    \begin{itemize}
    \item If there is no refreshment or of a register that does not
      contain $a$, then it is trivial;
    \item If there is a refreshment of the register containing $a$,
      then either we replace the new value by $b$ in $m'$ if $a$ is
      guessed again, or we keep the new value. Since $b\notin
      \text{codom}(m')$ this is a valid transition. Note that in this
      case $a\notin \text{codom}(m')$, and thus $m'=m'\delta_{a,b}$.
    \end{itemize}
  \item Otherwise if the transition reads the register containing $a$,
    then the same transition, after the replacement of $a$ by $b$,
    will read $x\delta_{a,b}=b$. Otherwise it is trivial.
  \end{itemize}
\end{proof}

By recursion on the length of a run and by changing the possible
occurrences of $b$ in the intermediate configurations (possible since
the letter $b$ is never read in a transition) we obtain

\begin{lemma}{\label{lem:word:replacement}}
  If $s,m \Longrightarrow^\omega s',m'$, $a\in \text{codom}(m)$ and $b
  \notin \text{codom}(m) \cup\text{codom}(m')$ and $\vert
  \omega\vert_b=0$ then $s,m\delta_{a,b}
  \Longrightarrow^{\omega\delta_{a,b}} s',m'\delta_{a,b}$
\end{lemma}

The following lemmas hold for any weakly deterministic NFMA
recognizing the language $\mathcal{L}_n$.

\begin{lemma}{\label{lem:active:variables}}
  In any successful run recogniznig a word in $L_\omega$:
  \begin{enumerate}
  \item Once a register $r$ is read during the initialization phase it
    is not refreshed before the pivot point;
  \item If $r_1,\ldots,r_n$ are the registers read at the end of the
    initialization phase, then the rest of the run reads no other
    registers ($\varepsilon$-transitions and reading the letter $\#$
    is still possible)
  \end{enumerate}
\end{lemma}

\begin{proof}
  The first point is trivial, as Lemma~\ref{lem:word:replacement}
  would otherwise permit us to construct another successful run in
  which the letter guessed is replaced by another one (not occurring
  in $\omega$), thereby contradicting the hypothesis that the NFMA
  recognizes exactly $\mathcal{L}_n$.

  The second point is a direct consequence of the first point, as
  other registers contain letters which do not occur in $\omega$, and
  thus which are not in any word belonging to $L_\omega$. But
  $L_\omega$ is exactly the subset of $\mathcal{L}_n$ of words
  beginning with $\omega$, so a successful run reading first $\omega$
  must recognize a word in $L_\omega$.
\end{proof}

As a consequence, the possible configurations of the automaton are
characterized by their values on the registers $r_1,\ldots,r_k$, and
these values are fixed at the end of the initialization phase.

\begin{theorem}
  A weakly deterministic NFMA recognizing $\mathcal{L}_n$ has a at
  least $n^n$ states.
\end{theorem}

\begin{proof}
  Let $\mathcal{A}$ be any weakly deterministic NFMA recognizing a
  language $\mathcal{L}_n$. 
  Let $\alpha\neq\beta$ be two words such that
  $\omega_\alpha=\omega\cdot\alpha\cdot\#\cdot\tilde\alpha\tilde\omega$ and 
  $\omega_\beta=\omega\cdot\beta\cdot\#\cdot\tilde\beta\tilde\omega$ are in
  $L_\omega$.

  Let $R_\alpha$ (resp. $R_\beta$) be a run recognizing
  $\omega_\alpha$ (resp. $\omega_\beta$), and $s_\alpha$
  (resp. $s_\beta$) be the pivot point in $R_\alpha$
  (resp. $R_\beta$).  Let $s^\alpha_n,m^\alpha_n$
  $s^\beta_n,m^\beta_n$ be a configuration reached after reading
  $\omega$ in each run. Since the automaton is weakly deterministic we
  have $s^\alpha_n=s^\beta_n$. By Lemma~\ref{lem:active:variables},
  point 1., the active variables in these states are the registers
  $r_1,\ldots,r_n$ storing the letters of $\omega$.

  Let $p_\alpha,m_\alpha$ (resp. $p_\beta,m_\beta$) be the pivot
  configuration in the run recognizing $\omega_\alpha$
  (resp. $\omega_\beta$). Again by Lemma~\ref{lem:active:variables},
  point 1., the active variables in these configurations are the
  registers $r_1,\ldots,r_n$ and they hold the same values in the two
  runs.
  
  Assume that $p_\alpha=p_\beta$. Then
  Lemma~\ref{lem:active:variables}, point 2, implies that the word
  $\omega\cdot\alpha\cdot\#\cdot \tilde\beta\cdot\tilde\omega$ is also
  recognized by $\mathcal{A}$, even though it is not in
  $\mathcal{L}_n$. Thus this contradicts the assumption the
  $\mathcal{A}$ recognizes $\mathcal{L}_n$.

  As a consequence, there is at least as many pivot states as the
  number of possible words $\alpha$ such that
  $\omega\cdot\alpha\cdot\#\cdot\tilde\alpha\cdot\tilde\omega\in\mathcal{L}_n$.
  Hence our lower bound $n^n$ on the number of pivot states, and thus
  on the size of any weakly deterministic NFMA recognizing
  $\mathcal{L}_n$.
\end{proof}

%% file: properties.tex
\section{Properties of  guarded variable automata}
\label{prop:sec}

We study the closure properties of GVAs and some basic decision problems. 
\ucomment{
Firstly, we show  that GVAs and GVAs with empty initial assignment  
recognize the same languages. 
The idea
is to turn the initial assignment $\bs{\tau_0}$  into the  guard  
$\phi = \bigland \set{(z=a) \gvert \bs{\tau_0}(z)=a}$
and propagate it in 
the transitions of the automaton,  and remove  from this guard the variables which are refreshed at 
each  state. 

\begin{lemma}
  For a  GVA (with initial assignment)
  we can construct a GVA  with empty initial assignment recognizing the same language.
\end{lemma}
}

Since GVAs and NFMA recognize the same languages, 
 GVAs inherit all the closure properties of NFMA. Hence,    

\begin{theorem}
  GVAs are closed under union, concatenation, Kleene operator and
  intersection. They are not closed under complementation.
\end{theorem}

\ucomment{\noindent 
For the closure under union  we simply take the disjoint union of the two GVAs.
The closure under Kleene operator  and concatenation follows from  [[]].
 The closure under intersection  is a consequence of the   fact that computing  the  
intersection of two GVAs amounts 
to computing their Cartesian product,  which can be   turned  into  a GVA 
(Appendix \ref{closure:sec:annex}).
For the complementation, consider the language $L_2$ of all the words in which 
there is  a letter  that occurs at least twice. In fact, $L_2$ is GVA-recognizable 
\cite{BCR:FVA:13}. The complement of $L_2$ consists of all the words
in which all the letters are different,
which is not GVA-recognizable since we need to compare each letter $w_j$ to all the previous letters
$w_i, i <j$. Thus we need an infinite number of states.     
}

Despite GVAs are not closed under complementation,
 FAs can be complemented within the class of GVAs. 
  That is, given a FA $F$ there exists a GVA $\mycal{A}$ such that
  $L(\mycal{A})= \Sigma^{\star} \setminus L(F)$,
    see Proposition \ref{complement:prop} in Appendix \ref{closure:sec:annex}.
It is worth mentioning that FAs cannot  be complemented within the subclass
of FVAs. 


We study the decidability and complexity of classical decision
problems: Nonemptiness (given $\mycal{A}$, is $L(\mycal{A})\neq
\emptyset$?), Membership (given a word $w$ and $\mycal{A}$, is $w \in
L(\mycal{A})$?), Universality (given $\mycal{A}$, is
$L(\mycal{A})=\Sigma^*$?), and Containment (given $\mycal{A}_1$ and
$\mycal{A}_2$, is $L(\mycal{A}_1) \subseteq L(\mycal{A}_2)$?). 


\begin{theorem}
  \label{univ:GVA:th:membership}
  For GVAs, Membership is NP-com\-plete,  
  Universality and Containment are  undecidable.
\end{theorem}
\LongVersion{
\begin{proof}
  Let $\mycal{A}$ be a GVA and $w=w_1 \cdots w_n$ a word in  $\Sigma^{\star}$. 
  For the upper  bound of Membership, a non deterministic polynomial algorithm 
  guesses a   path  in  $\mycal{A}$ of length $|w|$ such that the final state is accepting; 
  and a series of substitutions   $\sigma_1,\cdots,\sigma_{|w|}$, 
  then checks wether  the corresponding  run on $w$ is possible.
  The lower bound, i.e. the NP-hardness, follows from the fact that the 
  membership problem for GVAs without guards, i.e. FVAs, is NP-complete\cite[Theorem 3]{BCR:FVA:13}. 
  The undecidability of Universality  follows from \cite{GKS10} where 
  this problem is shown undecidable for the class of variable automata which is a subclass of GVAs. 
  \qed 
\end{proof}
}

  The undecidability of Containment and Universality  
  is a consequence of the undecidability of these problems for NFMA \cite{ijfcs/KaminskiZ10}.
  However, the decidability of Containment  if one of the GVAs is a finite automaton 
  results from the fact that the intersection of the languages in this case is regular since 
  the Cartesian product of a GVA and a FA yields a FA.   
  The  proof is the same 
  as that of  Lemma 17 of \cite{BCR:FVA:13}.   Hence, 
\begin{proposition}
The containment problems between a GVA and a FA are decidable.
\end{proposition}

We show in the next subsection  that 
nonemptiness for GVAs is PSPACE-Complete.


%% file: emptiness.tex
\subsection{Nonemptiness is  PSPACE-Complete}

We recall that Nonemptiness is NL-Complete for  both FVAs \cite{BCR:FVA:13} and 
variable automata \cite{GKS10}, and it is NP-Complete for FMAs \cite{SakamotoI00}.
Firstly we show that Nonemptiness is in PSPACE.
\ucomment{
\begin{definition}{(Configuration)}
  Let $\mathcal{A}$ be a GVA with $k$ variables over the alphabet
  $\Sigma$. A \emph{configuration} of $\mathcal{A}$ is a tuple
  $(q,\alpha_1,\ldots,\alpha_n)$ with
  $\alpha_1,\ldots,\alpha_n\in\Sigma$ and $q$ is a state of
  $\mathcal{A}$. We also say that $q$ is the state of
  $(q,\alpha_1,\ldots,\alpha_n)$, and that a configuration is initial
  (\textit{resp.} final) if its state is initial (\textit{resp.}
  final).
\end{definition}

\begin{definition}{(Run)}
  Let $\mathcal{A}$ be a GVA with $n$ variables over the alphabet
  $\Sigma$. A \emph{run} of $\mathcal{A}$ is a sequence of
  configurations $m_1,\ldots,m_l$ such that $m_1$ is an initial
  configuration and, denoting
  $m_i=(q_i,\alpha_1^i,\ldots,\alpha_n^i)$, there exists a letter
  $\beta_i$ such that $\delta_{\mathcal{A}}(m_i,\beta_i,m_{i+1})$.
\end{definition}
}
Given a GVA $\mycal{A}$, we shall show that $\mycal{A}$ recognizes 
a non-empty language over $\Sigma$ iff  
$\mycal{A}$ recognizes a non-empty language over a \emph{finite} set of letters. 
For this purpose, and in order to relate the two runs of $\mycal{A}$ 
(the one over an infinite alphabet  and the  one over a finite alphabet) 
we introduce the relation of \emph{coherence} between 
substitutions.  

\begin{definition}
  \label{coherence:subs:def}
  Let $C$ be a  finite subset of  $\Sigma$. 
  The coherence relation $\synch_{C} \subseteq \zeta \times \zeta$ between 
  substitutions 
  is defined by 
  $\bar{\sigma} \synch_{_{C}} \sigma$  iff the three following conditions hold:
  \begin{enumerate}
  \item $dom(\bar{\sigma}) = dom(\sigma)$,
  \item If $\bar{\sigma}(x) \in C$ then  $\bar{\sigma}(x)=\sigma(x) $, and 
    if  $\sigma(x) \in C$, then $\bar{\sigma}(x)=\sigma(x) $, for any variable $x \in dom(\sigma)$, and 
  \item for any variables $x,y \in  dom(\sigma)$,  $\bar{\sigma}(x)=\bar{\sigma}(y)$ iff $\sigma(x)=\sigma(y)$.
  \end{enumerate} 
\end{definition}

We need to define a function $\bs{\Theta}$ that will be used in the proof of  Lemma \ref{lemma:finite:constants} 
and other Lemmas.
Given two
sets $S_1$ and $S_2$ of letters  such that $S_1 \cap S_2=C \neq
\emptyset$,  $|S_1\setminus S_2| \ge |\mycal{X}|$ and $|S_2 \setminus S_1| \ge |\mycal{X}|$, 
we can define a function:    
  \begin{align}
  \label{Theta:def:eq}
    \bs{\Theta}_{C}^{S_1,S_2} &: \xi_{\mycal{X},S_1}  \times \xi_{\mycal{X},S_1} \times \xi_{\mycal{X},S_2}  \uberTo{} \xi_{\mycal{X},S_2}
  \end{align}
such that given three substitutions $\sigma,\gamma,\sigma'$, such that $dom(\sigma)\cap dom(\gamma)=\emptyset$
and $\gamma \synch_C \gamma'$  construct a substitution  $\gamma'=\bs{\Theta}_{C}^{S_1,S_2}(\sigma,\gamma,\sigma')$
such that $\sigma \uplus \gamma \synch_C \sigma' \uplus \gamma'$.

\begin{lemma}
 \label{lemma:finite:constants}
  Let $\mathcal{A}$ be a GVA  over $\Sigma$ with $k$ variables and $m$ constants
  $\Sigma_{\mycal{A}}=\set{c_1,\ldots,c_m}$. Let  $\bs{\Sigma}=\set{a_1,\ldots,a_k,c_1,\ldots,c_m}$. 
  Then, $\mycal{A}$ recognizes a non-empty language  over
  $\Sigma^\star$ if, and only if, it recognizes a non-empty language
  over $\bs{\Sigma}^{\star}$.
\end{lemma}
\begin{proof} (Sketch)
  Let $C=\set{c_1,\ldots,c_k}$.
  We show that there is a run 
$(\sigma_0,q_0)\uberTo{} \ldots \uberTo{}(\sigma_n,q_n)$  over  $\Sigma^{\star}$ in $\mycal{A}$ 
  iff  there is a run 
 $(\sigma'_0,q_0)\uberTo{} \ldots \uberTo{}(\sigma'_n,q_n)$ over   
 $\bs{\Sigma}^{\star}$ in $\mycal{A}$
  such that $\sigma_i \synch_{C} \sigma'_i$, for all $i=0,\ldots,n$. 
The proof is by induction
on $n$ in both directions. The base case $n=0$ holds trivially since $\sigma_0=\sigma'_0=\emptyset$. Assume that 
the claim holds up to $n$ and let us prove it for $n+1$.
\begin{itemize}
\item[{$\Rightarrow$)}] 
  Assume there is a transition $q_n \uberTo{\alpha_n,g_n} q_{n+1}$ in $\mycal{A}$ where $\alpha_n \in \Sigma \cup \mycal{X}$ and 
  $g_n$ is a guard.  From the induction hypothesis we have that $\sigma_n \synch_{C} \sigma'_n$. It follows that  
   that $\sigma_{n}(g_n)$ holds iff $\sigma'_{n}(g_n)$ holds (Lemma \ref{synch:subst:guards:lemma} in Appendix \ref{req:sec:annex}). 
  Thus, the transition in $\mycal{A}$ over $\bs{\Sigma}$ is possible.  We describe next this transition.  
  From Definition \ref{run:lang:def} of the run of  GVAs, there 
  exists a substitution 
  $\gamma_n: \mycal{V}(\sigma_n(\alpha_n)) \cup \mycal{V}(\sigma_n(g_n)) \uberTo{} \Sigma$ such that 
   $(\gamma_n \uplus \sigma_n)(g_n)$ holds.
  Hence, we must find a substitution 
  $\gamma'_n: \mycal{V}(\sigma'_n(\alpha_n)) \cup \mycal{V}(\sigma'_n(g_n)) \uberTo{} \bs{\Sigma}$ such that 
   $(\gamma'_n \uplus \sigma'_n)(g_n)$ holds and $\sigma_{n+1}\synch_{C} \sigma'_{n+1}$. 
 We define $\gamma'_n$ by $\gamma'_n= \bs{\Theta}_{C}^{\Sigma,\bs{\Sigma}}(\sigma_n,\gamma_n,\sigma'_n)$.

\item[$\Leftarrow$)] Same proof but we call the function $\bs{\Theta}_{C}^{\bs{\Sigma},\Sigma}(\sigma'_n,\gamma'_n,\sigma_n)$.
\end{itemize}
\qed
\end{proof}

\begin{definition}{(Restricted configuration)}
  A \emph{restricted configuration} of a GVA $\mathcal{A}$ with $k$ variables
  $x_1,\ldots,x_k$ and $m$ constants $c_1,\ldots,c_m$ is a tuple
  $(q,\alpha_1,\ldots,\alpha_k)$ where $\alpha_1,\ldots,\alpha_k \in
  \left\lbrace a_1,\ldots,a_k,c_1,\ldots,c_m \right\rbrace$.
\end{definition}

Note that the number of different
restricted configurations is exponentially  bounded in the size of $\mathcal{A}$.

\begin{lemma}{\label{lemma:different:restricted:configurations}}
  Assume $\mathcal{A}$ is a GVA that recognizes a non-empty
  language. Then there is an accepting  run 
  $\bs{q}_1\uberTo{}\ldots \uberTo{} \bs{q}_l$ such that $i\neq j$ implies $\bs{q}_i\neq  \bs{q}_j$, 
  and each $\bs{q}_i$ is a restricted configuration.
\end{lemma}
\begin{proof}
  By Lemma~\ref{lemma:finite:constants} and the assumption that
  $\mathcal{A}$ recognizes a non-empty language, it recognizes a word
  in $\left\lbrace a_1,\ldots,a_k,c_1,\ldots,c_m
  \right\rbrace^\star$. Let $\omega$ be such a word of minimal length,
  and let $\bs{q}_1 \uberTo{}\ldots \uberTo{} \bs{q}_l$ be an accepting  run of $\mathcal{A}$ recognizing
  $\omega$. 
  Since all the variables are instantiated with constants in
  $\left\lbrace a_1,\ldots,a_k,c_1,\ldots,c_m \right\rbrace$, each
  $m_i$ is a restricted configuration of $\mathcal{A}$. 
  Furthermore, if a restricted configuration appears at steps $i$ and
  $j$ with $i\neq j$, it can be shown that
  $\bs{q}_1 \uberTo{}\ldots\uberTo{}\bs{q}_i\uberTo{}\bs{q}_{j+1}\uberTo{}\ldots \uberTo{} \bs{q}_l$
  is also a run for the automaton
  $\mathcal{A}$ that contradicts the minimality of $\omega$. Hence for
  $i\neq j$ we have $\bs{q}_i\neq \bs{q}_j$.
  \qed
\end{proof}

As a corollary, we obtain that if a GVA recognizes a non-empty
language $L$ it has an accepting run consisting of restricted
configurations that each appear at most once. Hence its length is less
than the number of configurations, which can be encoded in binary in
space linear in the size of $\mathcal{A}$. We thus obtain: 

\begin{theorem}
\label{theorem:emptyness:in:PSPACE}
  The nonemptiness  problem for GVAs is in  PSPACE.
\end{theorem}

As a direct consequence of Subsection \ref{subsec:gva:nfma:equiv}, we also get:
\begin{corollary}
\label{corollary:emptyness:in:PSPACE}
  The nonemptiness  problem for NFMAs is in  PSPACE.
\end{corollary}

To show that Nonemptiness of GVAs is PSPACE-hard, 
we reduce the reachability problem for bounded one-counter
automata (known to be PSPACE-hard) to the nonemptiness problem of GVAs. 
In the rest of this
section, we first present bounded one-counter automata, and then
proceed to GVAs. 


\begin{definition}{(Bounded one-counter automata~\cite{counterautomatadef})}
  A \emph{bounded one-counter automaton} (\ttt{Boca}) is a tuple
  $(Q,b,\Delta,q_0)$ where $Q$ is a finite set of states, $b\in \N$ is
  a global counter bound, $q_0$ is the initial state, and
  $\Delta\subset Q\times [-b,b] \times Q$ is the transition relation.

  A \ttt{Boca} configuration is a tuple $(q,p)$ with $q\in Q$ and $0\le p
  \le b$. Given $\tau=(q,p,q')\in\Delta$ and two configurations
  $c_1=(q_1,p_1)$ and $c_2=(q_2,p_2)$ we denote $c_1\to^\tau c_2$ if 
  $q_1=q$, $q_2=q'$,  $p_2=p_1+p$.
\end{definition}
Note that mandating that $c_2$ is a configuration implies two implicit
inequality testing $0\le p_1+p\le b$. The reachability problem for
\ttt{Boca} consists in determining whether there exists a sequence of
transitions starting from the configuration $(q_0,0)$ and ending in
a configuration $(q,p)$. This problem has been shown to be 
PSPACE-complete in~\cite{counterautomatapspace}. We reduce it to GVA's Nonemptiness  as follows:                                 
\begin{itemize}
\item Assuming $2^{k-1}\le b \le 2^{k}$, we build a GVA with $k+1$
  variables $x_1,\ldots,x_k,r$, where each $x_i$ contains the $i^{\rm{th}}$
  bit in the binary representation of the counter $p$, and $r$ stands
  for a register that will hold the current carry of a bit-per-bit
  binary addition;
\item Addition of a positive constant to the counter is encoded by
  a sequence of $k+1$ half-adders. Each half-adder reads $x_i$,
  and depending on the value of $r$ and the $i^{\rm{th}}$ bit of the number to
  add, refreshes $x_i$ and $r$, and sets (using a guard) the correct
  new value for these two variables. A constant number of states each
  necessary at each half-addition step, and thus the encoding is
  linear in $k=\ceil{\log_2 b}$, and thus in the size of the input
  \ttt{Boca}. Note that the result is greater than $2^{\ceil{\log_2 b}}$
  if, and only if, the register $r$ contains $1$ in the end;
\item Bitwise 2-complement of a register is similarly performed
  bit-per-bit, using a number of states linear in $k$;
\item Substraction of a positive integer is encoded by two bitwise
  2-complement computations on the counter around a positive integer
  addition. If the final counter is negative the intermediate addition
  step will yield a number greater than $2^k$;
\item Finally, we test after each addition of a positive integer that
  the counter is smaller or equal to $b$ by adding $2^k-b$ to it, and
  then substracting this same number;
\item The final configuration $(q_f,p_f)$ is encoded by a transition
  from states of the GVA encoding $q_f$ to its unique final state
  having as guard the equality test of the variables $x_1,\ldots,x_k$
  with the bits in the binary encoding of $p_f$.
\end{itemize}
 The PSPACE-hardness of the reachability problem for \ttt{Boca}
 and  Theorem \ref{theorem:emptyness:in:PSPACE} imply:

\begin{theorem}
  The nonemptiness problem for GVAs is PSPACE-complete. 
\end{theorem}

Note that we need only 
two letters in $\Sigma_{\mycal{A}}$ 
in the encoding employed to prove the hardness.


%% file: simulation.tex
\section{Simulations for   GVAs}
\label{simulation:sec}

We define and study the  simulation preorder for GVAs, an extension  of the  simulation preorder for FAs.  
To simplify the presentation, we shall only consider in this section
GVAs without $\eps$-transitions and in which there is a unique initial state
and all the states are accepting.
The definition of simulation preorder for GVAs follows.

\begin{definition}
  \label{sim:GVA:def}
  Let   $\mycal{A}_1=\model{\Sigma,\mathcal{X}_1,Q_1,q^{1}_0,\delta_1,F_1,\kappa_1}$
  and  $\mycal{A}_2=\model{\Sigma,\mathcal{X}_2, Q_2,q^2_0,\delta_2,F_2,\kappa_2}$ be two GVAs where
  $\mycal{X}_1 \cap \mycal{X}_2=\emptyset$.
  A simulation of  $\mycal{A}_1$ by $\mycal{A}_2$ is a relation 
  $\unlhd  \subseteq (\zeta_{\mycal{X}_1,\Sigma} \times Q_1) \times (\zeta_{\mycal{X}_2,\Sigma} \times Q_2)$ such that:
  \begin{itemize}
  \item $(\emptyset,q^1_0) \unlhd (\emptyset,q^2_0)$.
  \item if $(\sigma_1,q_1) \unlhd (\sigma_2,q_2)$ and if $(\sigma_1,q_1) \uberTo{a} (\sigma'_1,q'_1)$ for $a\in \Sigma$ 
    then there exists a state $q'_2 \in Q_2$ and a substitution $\sigma'_2$ 
    such that   $(\sigma_2,q_2) \uberTo{a} (\sigma'_2,q'_2)$ and 
    $(\sigma'_1,q'_1) \unlhd (\sigma'_2,q'_2)$. 
  \end{itemize}
\end{definition}

In  order to study the decidability of the simulation,
we provide next an equivalent  game-theoretic definition  
in which  we make explicit the evolution of the configurations.

\begin{definition}
  \label{sim:game:def}
  Let
  $\mycal{A}_1=\model{\Sigma,\mathcal{X}_1,Q_1,q^{1}_0,\delta_1,F_1,\kappa_1}$
  and
  $\mycal{A}_2=\model{\Sigma,\mathcal{X}_2,$ $Q_2, q^2_0,\delta_2,F_2,\kappa_2}$
  be two GVAs where $\mycal{X}_1 \cap \mycal{X}_2=\emptyset$.  
  Let $\Pos$: 
  \begin{align*}
     \Pos \; \subseteq  \; &  (\zeta_{\mycal{X}_1} \times Q_1)
    \times (\zeta_{\mycal{X}_2} \times Q_2)   \;\; \cup \\
    & (\zeta_{\mycal{X}_1} \times Q_1) \times
    (\zeta_{\mycal{X}_2} \times Q_2) \times (\zeta_{\mycal{X}_1} \times   (\mcons)) 
  \end{align*}
  be the set of positions reachable from
  $p^{\star}=\AbelardM{(\emptyset,q^1_0),(\emptyset,q^2_0)}$ by the set
  of moves $M=M_{A} \cup  M_E$, where:
  $$
  \begin{array}{rl}
    M_A&=\big\{ \AbelardM{(\sigma_1,q_1), \tau_2} \uberTo{}  
    \EloiseM{((\sigma_1 \uplus \gamma)_{|D},q'_1),  \tau_2}{(  \sigma_1 \uplus \gamma, \alpha)} \\ 
    &\hspace*{1cm}\vert\hspace*{0.5cm}  q'_1 \in \delta_1(q_1,\alpha,g_1)\\
    &\hspace*{1.5cm} \tand D=Dom(\sigma_1 \uplus \gamma) \setminus \kappa_1^{-1}(q'_1)\\  
    &\hspace*{1.5cm} \tand   \sigma_1 \uplus \gamma \models g_1 \\ 
    &\hspace*{1.5cm} \tand \gamma: \mycal{V}(\sigma_1(\alpha)) \cup \mycal{V}(\sigma_1(g_1)) \to \Sigma \big\} \\ 
    M_E &=\big\{ \EloiseM{(\sigma_1,q_1),(\sigma_2,q_2)}{(\sigma_3,\alpha)}
    \uberTo{} \AbelardM{(\sigma_1 ,q_1), ((\sigma_2 \uplus \gamma_2)_{|D_2},q'_2)}\\
    &\hspace*{1cm}\vert\hspace*{0.5cm}  q'_2 \in \delta_2(q_2, \beta,g_2)\\
    &\hspace*{1.5cm} \tand D_2=Dom(\sigma_2 \uplus \gamma_2) \setminus \kappa_2^{-1}(q'_2)\\
    &\hspace*{1.5cm} \tand \gamma_1(\sigma_3(\alpha))=\gamma_2(\sigma_2(\beta)) \\
    &\hspace*{1.5cm} \tand    \gamma_2 \models  \sigma_2(g_2) \\ 
    &\hspace*{1.5cm} \tand \gamma_2: \mycal{V}(\sigma_2(\beta)) \cup \mycal{V}(\sigma_2(g_2)) \to \Sigma \big\}\\
  \end{array}
  $$
  with $\tau_2$ a  configuration  in 
  $\zeta_{\mycal{X}_2 \times \Sigma} \times Q_2$, and $\sigma_1,\sigma_2,\sigma_3$ are substitutions.   
  We let $\Pos_E=\Pos\cap (\zeta_{\mycal{X}_1} \times Q_1) \times
  (\zeta_{\mycal{X}_2} \times Q_2) \times (\zeta_{\mycal{X}_1} \times
  \mcons)$ and $\Pos_A = \Pos \cap (\zeta_{\mycal{X}_1} \times Q_1)
  \times (\zeta_{\mycal{X}_2} \times Q_2)$.  The \emph{simulation game} of
  $\mycal{A}_1$ by $\mycal{A}_2$, denoted by
  $\mycal{G}(\mycal{A}_1,\mycal{A}_2)$, is the two-players game $\langle
  \Pos_{E},\Pos_{A},M, p^{\star} \rangle$. As usual, any infinite play is
  winning for \Eloise, and any finite play is losing for the player who
  cannot move. And thus we  write $\mycal{A}_1 \unlhd \mycal{A}_2$.
\end{definition}

\emph{The simulation problem} for GVAs is the following: 
given two GVAs  $\mycal{A}_1$ and  $\mycal{A}_2$,  is $\mycal{A}_1 \unlhd \mycal{A}_2$?

\section{Decidability of the simulation problem}
\label{decidability:sec}
In this section  we  show that  the  {simulation}  problem  is decidable.
The idea is that this problem can be reduced to a {simulation} problem 
over the same  GVAs in which  the two players instantiate the variables  from a \emph{finite} set
of letters, as proven in Proposition \ref{sigma:finite:prop}.

\begin{definition}
  \label{sigma:finite:strategy:def}
  Let
  $\mycal{A}_1=\model{\Sigma,\mathcal{X}_1,Q_1,q^{1}_0,\delta_1,F_1,\kappa_1}$
  and $\mycal{A}_2=\langle\Sigma,\mathcal{X}_2,$ $Q_2,  q^2_0, \delta_2,\\ \
  F_2,\kappa_2\rangle$ be two GVAs. Let  $k=|\mycal{X}_1|+|\mycal{X}_2|$.  We define
  $\overline{\mycal{G}}(\mycal{A}_1,\mycal{A}_2)$ to be the game
  obtained by restricting the codomain of $\gamma$ to $C_0$ in the rules
  of \Eloise $M_{E}$ and \Abelard $M_{A}$ in Def.
  \ref{sim:game:def}, where 
  \begin{align}
    \label{C0:Def}  
      C_0=(\Sigma_{\mycal{A}_1} \cup  \Sigma_{\mycal{A}_2}) \uplus \set{c_1,\ldots,c_k}
  \end{align}
\end{definition}
The following Lemma states an immediate   property of the game $\overline{\mycal{G}}$.
\begin{lemma}
  \label{Gbar:finite:lemma}
  Let $\mycal{A}_1,\mycal{A}_2$ be two GVAs.
  Then, the game $\overline{\mathcal{G}}(\mycal{A}_1,\mycal{A}_2)$ is
  finite.
\end{lemma}

\ucomment{\noindent In order to prove  Proposition \ref{sigma:finite:prop} we need
  to introduce the notion of coherence between substitutions and 
  between  game positions. The coherence relation was introduced in
  \cite{BCR:FVA:13}, we reproduce it here.

  \begin{definition}
    \label{coherence:subs:def}
    Let $C$ be a  finite subset of  $\Sigma$. 
    The coherence relation $\synch_{C} \subseteq \zeta \times \zeta$ between 
    substitutions 
    is defined by 
    $\bar{\sigma} \synch_{_{C}} \sigma$  iff the three following conditions hold:
    \begin{enumerate}
    \item $dom(\bar{\sigma}) = dom(\sigma)$,
    \item If $\bar{\sigma}(x) \in C$ then  $\bar{\sigma}(x)=\sigma(x) $, and 
      if  $\sigma(x) \in C$, then $\bar{\sigma}(x)=\sigma(x) $, for any variable $x \in dom(\sigma)$, and 
    \item for any variables $x,y \in  dom(\sigma)$,  $\bar{\sigma}(x)=\bar{\sigma}(y)$ iff $\sigma(x)=\sigma(y)$.
    \end{enumerate} 
  \end{definition}
}

\noindent 
In order to prove  Proposition \ref{sigma:finite:prop} we need
to adapt  the notion of coherence between substitutions given in Definition \ref{coherence:subs:def} 
to the  coherence between  game positions.
The definition of the coherence between game positions, still denoted by $\synch_C$,
follows.

\begin{definition}
  \label{coherence:posi:def}
  Let $C $ be a  finite subset of  $\Sigma$, and
  $\mycal{A}_1=\langle\Sigma,\mathcal{X}_1,Q_1,q^{1}_0,\delta_1,F_1,\kappa_1\rangle$,
  and
  $\mycal{A}_2=\model{\Sigma,\mathcal{X}_2,Q_2,q^2_0,\delta_2,F_2,\kappa_2}$
  be two GVAs s.t. $\mycal{X}_1$ $\cap \mycal{X}_2=\emptyset$.  
  Let
  $\Pos_E$ (resp. $\Pos_A$) be the set of \Eloise's (resp. \Abelard's)
  positions in the game $\mycal{G}(\mycal{A}_1,\mycal{A}_2)$.  Then we
  define the relation: $ \synch_{C} \; \subseteq \Pos_A\times \Pos_A
  \,\cup \, \Pos_E\times \Pos_E $
  by:  
  \begin{description} 
  \item[$\bullet$] For any substitutions  $\sigma_i,\bar\sigma_i$ of proper domain
    included in $\mycal{X}_i$ ($i=1,2$) we have:\\
    $\big( \AbelardM{(\bar{\sigma}_1,q_1),(\bar{\sigma}_2,q_2)} \synch_{C} \AbelardM{(\sigma_1,q_1), 
      (\sigma_2,q_2)}\big)$ iff 
    $   (\bar{\sigma}_1 \uplus \bar{\sigma}_2) \synch_{C}  (\sigma_1 \uplus \sigma_2)$.
  \item[$\bullet$] For any $\sigma_i,\bar\sigma_i$ of proper domain
    included in $\mycal{X}_i$ ($i=1,2$), for any  substitutions
    $\sigma,\bar\sigma$ with  proper domain included in 
    $\mycal{X}_1$, we have: 
    $((\bar{\sigma}_1 \cup \bar{\sigma}) \uplus \bar{\sigma}_2 )
    \synch_{C} (({\sigma}_1\cup {\sigma}) \uplus {\sigma}_2 )$ iff 
    $\big(
    \EloiseM{(\bar{\sigma}_1,q_1),(\bar{\sigma}_2,q_2)}{(\bar{\sigma},\alpha)}
    \synch_{C} \EloiseM{(\sigma_1,q_1), (\sigma_2,q_2)}{(\sigma,\alpha)}\big)$.
  \end{description} 
\end{definition}

In order to prove Proposition \ref{sigma:finite:prop}, we need 
to prove a technical Lemma:

\begin{lemma}
  \label{f:construct:simul:lemma}
  Let $\mycal{A}_1=\model{\Sigma_1,\mathcal{X}_1,Q_1,Q_0^1,\delta_1,F_1,\kappa_1}$ and 
  $\mycal{A}_2=\model{\Sigma_2,\mathcal{X}_2,Q_2,Q_0^2,\delta_2,F_2,\kappa_2}$ be two GVAs, and let 
  $\mycal{X}=\mycal{X}_1 \cup \mycal{X}_2$.
  Let $\mycal{G}(\mycal{A}_1,\mycal{A}_2)$ (resp.  
  $\overline{\mycal{G}}(\mycal{A}_1,\mycal{A}_2)$)  be the  simulation game in which the two players 
  instantiate the variables from $\Sigma$ (resp. $C_0$ defined  in Eq. (\ref{C0:Def})).
  Let $\wp^\star$ and $\overline{\wp}^{\star}$ be their starting position respectively.
  Then, there is a   function $f: \, Pos(\mycal{G}(\mycal{A}_1,\mycal{A}_2)) \uberTo{} Pos(\overline{\mycal{G}}(\mycal{A}_1,\mycal{A}_2))$
  with $f(\wp^\star)=\overline{\wp}^{\star}$ and $\wp \synch_{C} f(\wp)$ for all $\wp \in Pos(\mycal{G}(\mycal{A}_1,\mycal{A}_2)$,
  such that the following hold:  
  \begin{enumerate}[i)]
  \item  for all $\overline{\wp} \in  \Pos_{A}(\overline{\mycal{G}}(\mycal{A}_1,\mycal{A}_2))$, 
    if $\overline{\wp} \uberTo{} \overline{\wp}'$ is a move of \Abelard in $\overline{\mycal{G}}(\mycal{A}_1,\mycal{A}_2)$ 
    and $f(\wp)=\overline{\wp}$ for some  position $\wp$ in $\mycal{G}(\mycal{A}_1,\mycal{A}_2)$, 
    then there exists a position $\wp'$ in $\mycal{G}(\mycal{A}_1,\mycal{A}_2)$  such that the move 
    $\wp \uberTo{} \wp'$ is possible in $\mycal{G}(\mycal{A}_1,\mycal{A}_2)$ and $f(\wp')=\overline{\wp}'$. And, 
  \item for all $\wp \in  \Pos_{E}(\mycal{G}(\mycal{A}_1,\mycal{A}_2))$, 
    if $\wp \uberTo{} \wp'$ is a move of \Eloise  in $\mycal{G}(\mycal{A}_1,\mycal{A}_2)$ 
    then there exists a position $\overline{\wp}'$ in $\overline{\mycal{G}}(\mycal{A}_1,\mycal{A}_2)$  such that the move 
    $f(\wp) \uberTo{} \overline{\wp}'$ is possible in $\overline{\mycal{G}}(\mycal{A}_1,\mycal{A}_2)$ and $f(\wp')=\overline{\wp}'$.
  \end{enumerate}
\end{lemma}
\begin{proof}(Sketch)
  The main part of the proof consists in finding the right way to relate
  the instantiation of the variables in
  $\overline{\mycal{G}}(\mycal{A}_1,\mycal{A}_2)$ and
  $\mycal{G}(\mycal{A}_1,\mycal{A}_2)$.  
  \begin{enumerate}[i)]
  \item The function  $\bs{\Theta}_{C}^{S_1,S_2}$  of Eq (\ref{Theta:def:eq})
    allows to construct  the instantiation of the variables   by \Abelard in $\mycal{G}$
    out of the instantiation of variables by \Abelard in $\overline{\mycal{G}}$ as follows. 
    Assume
    $\bar{\wp}=((\bar{\sigma}_1,q_1),(\bar{\sigma}_2,q_2))_{A}$ is a position in
    $\overline{\mycal{G}}(\mycal{A}_1,\mycal{A}_2)$, and $\bar{\gamma}$ is an
    instantiation made by \Abelard from $\bar{\wp}$ (i.e. $\gamma$ in the move
    $M_A$ of Def. \ref{sim:game:def}), 
    and  that $\wp=((\sigma_1,q_1),(\sigma_2,q_2))_{A}$ is a position 
    in  $\mycal{G}(\mycal{A}_1,\mycal{A}_2)$ such that $f(\wp)=\bar{\wp}$.
    Then,  \Abelard's instantiation
    $\gamma$   from $\wp$ is  
    defined by 
    $\gamma=\bs{\Theta}_{C}^{C_0,\Sigma}(\bar{\sigma}_1\uplus \bar{\sigma}_2,\bar{\gamma}, \sigma_1\uplus   \sigma_2)$.
  \item We define  a function   $\Xi_{C}^{S_1,S_2}$ which is similar  to the function  $\bs{\Theta}_{C}^{S_1,S_2}$
    but allows to construct the instantiation of the variables by  \Eloise in $\overline{\mycal{G}}$
    out of the instantiation of variables by \Eloise in $\mycal{G}$.    \qed
  \end{enumerate}
\end{proof}
\noindent Now we are ready to show that the games $\mycal{G}$ and $\overline{\mycal{G}}$ are equivalent. 
\LongVersion{We extend the proof of \cite{BCR:FVA:13}. 
  The diffrence is that in \cite{BCR:FVA:13} 
  the player    instantiates only one variable, the one of the sent message. 
  However, here the player  instantiates  many variables, i.e.  the variable of the sent message and 
  the free variables of the guard.   }
We recall that the variables in
$\overline{\mycal{G}}(\mycal{A}_1,\mycal{A}_2)$ are instantiated from
the finite set of letters $C_0= (\Sigma_{\mycal{A}_1} \cup \Sigma_{\mycal{A}_2}) \uplus \set{c_1,\ldots,c_k},$
\textrm{ where } $k=|\mycal{X}_1| + |\mycal{X}_2|$.

For the direction "$\Rightarrow$" we show that out of a winning
strategy of \Eloise in ${\mycal{G}}(\mycal{A}_1,\mycal{A}_2)$ we
construct a winning strategy for her in
$\overline{\mycal{G}}(\mycal{A}_1,\mycal{A}_2)$.

\hspace{-0.55cm}
\begin{minipage}{.55\textwidth}
  For this purpose, we
  show that each move of \Abelard in
  $\overline{\mycal{G}}(\mycal{A}_1,\mycal{A}_2)$ can be mapped to an
  \Abelard move in $\mycal{G}(\mycal{A}_1,\mycal{A}_2)$, and that
  \Eloise response in $\mycal{G}(\mycal{A}_1,\mycal{A}_2)$ can be
  actually mapped to an \Eloise move in
  $\overline{\mycal{G}}(\mycal{A}_1,\mycal{A}_2)$.  
  Formally, we need to  define  a  function 
  $f: \, Pos(\mycal{G}(\mycal{A}_1,\mycal{A}_2)) \uberTo{} Pos(\overline{\mycal{G}}(\mycal{A}_1,\mycal{A}_2))$
  in order to  make possible  this mapping as shown in  the   Diagram on the right.  
  It follows that  this is sufficient  to argue that if there is an  infinite play in $\mycal{G}$
  then we can construct an infinite play in $\overline{\mycal{G}}$ 
\end{minipage}
\begin{minipage}{.45\textwidth}
\vspace{-0.95cm}
  \centering
  \vspace{0.2cm}
  \begin{tikzpicture}
    \matrix (m) [matrix of math nodes,row sep=3.5em,column sep=5.5em,minimum width=2em]
            {   \wp  & \overline{\wp}=f(\wp)  \\
              \wp'  & \overline{\wp}'=f(\wp')  \\
              \wp''  & \overline{\wp}''=f(\wp'')\\};
            \path [-stealth]
            (m-1-1) edge [dashed,->] node [above] {$f$} (m-1-2)
            
            (m-2-1) edge [dashed,->] node [above] {$f$} (m-2-2)
            
            (m-1-2) edge node [right] {\ttt{A}} (m-2-2)
            edge [dashed,-] (m-1-1)

            (m-3-1) edge [dashed,->] node [above] {$f$} (m-3-2)

            (m-2-2) edge node [right] {\ttt{E}} (m-3-2)

            (m-1-1) edge node [right] {\ttt{A}} (m-2-1)
            (m-2-1) edge node [right] {\ttt{E}} (m-3-1)

            -- (-1.65,2.2) node {$\mycal{G}(\mycal{A}_1,\mycal{A}_2)$}
            -- (1.5,2.2) node {$\overline{\mycal{G}}(\mycal{A}_1,\mycal{A}_2)$}; 
            \draw [thick, <-,color=blue] (-1.85,1.1) arc  (120:60:82pt);
            \draw [thick, ->,color=blue] (-1.85,-1.2) arc  (120:60:82pt);
  \end{tikzpicture}
\end{minipage}
as well. 
  We show in Lemma \ref{f:construct:simul:lemma} that it is possible to
  construct the function $f$. 
  The proof of the direction ($\Leftarrow$) is similar to the one of  ($\Rightarrow$), 
we follow the  same construction. 
  Therefore,

\begin{proposition}
  \label{sigma:finite:prop}
  Let $\mycal{A}_1$ and  $\mycal{A}_2$ be two GVAs. 
  Then, \Eloise has a 
  winning strategy in $\mathcal{G}(\mycal{A}_1,\mycal{A}_2)$ iff she 
  has a  winning strategy in $\overline{\mathcal{G}}(\mycal{A}_1,\mycal{A}_2)$.
\end{proposition}

It follows from  Lemma \ref{Gbar:finite:lemma} and Proposition \ref{sigma:finite:prop}: 
\begin{theorem}
  \label{main:decidable:th}
  The  {simulation} problem   for  GVAs is decidable.
\end{theorem}

Note that the simulation problem for GVAs is in APSPACE: A position of 
$\overline{\mycal{G}}(\mycal{A}_1,\mycal{A}_2)$, 
has size linear in the size of $\mycal{A}_1$
and $\mycal{A}_2$, in order to encode the substitutions and states. 
Hence the number of different  
positions   of $\overline{\mycal{G}}(\mycal{A}_1,\mycal{A}_2)$
is bounded exponentially in the size of $\mycal{A}_1$ and $\mycal{A}_2$.
From this we can deduce that a  polynomial space alternating Turing machine can  solve the simulation game:
universal states correspond to \Abelard  positions and 
existential states correspond to \Eloise positions. Hence, 

\begin{theorem}
  \label{simulation:decidable:th}
  The  {simulation} problem   for  GVAs is in EXPTIME.
\end{theorem}


%% file: synthesis.tex
\section{Application to service  composition} 
\label{service:synthesis}

\input example.tex

In order to solve the PROM composition problem we 
define next the asynchronous product of GVAs which generalizes 
the asynchronous product of FAs as given in \cite{MW08}.   

\begin{definition}
Given $n$  GVAs 
$\mathcal{A}_i=\model{\Sigma_i,\mathcal{X}_i,Q_i,Q^i_0,\delta_i,F_i,\kappa_i}$, 
their  asynchronous product $\mathcal{A}_1 \otimes \cdots \otimes \mathcal{A}_n$  
is a GVA:  $\model{\Sigma,\mathcal{X},Q,Q_0,\delta,F,\kappa}$, 
where\footnote{Up to variable  renaming, 
 we assume that $\mycal{X}_i\cap \mycal{X}_j=\emptyset$, for all $i\neq j$.}:     
\begin{description}             
\item[$\bullet$] $\Sigma=\cup_{i=1,\ldots,n}\Sigma_i$, $\mathcal{X}=\cup_{i=1        ,\ldots,n}\mathcal{X}_i$, 
\item[$\bullet$] $Q=Q_1\times \cdots \times Q_n$,
       $Q_0=Q^1_0 \times \cdots \times Q_0^n$, $F=F_1\times \cdots \times F_n$, 
 \item[$\bullet$]  $\delta$ is defined by:  
 $\boldsymbol{q} \in \delta(\boldsymbol{p},t)$ iff for some $i$, $\pi_i(\boldsymbol{q}) \in \delta_i(\pi_i(\boldsymbol{p}),t)$, and for 
 all $j\neq i$ we have that $\pi_j(\boldsymbol{q})=\pi_j(\boldsymbol{p})$, where $\pi_i$ denotes the  projection
 along the  $i^{th}$-component, and
 \item[$\bullet$] $\kappa$ is defined by:  
  $ \boldsymbol{p} \in \kappa(x) $ iff for some $i$, $\pi_i(\boldsymbol{p}) \in \kappa_i(x)$.
\end{description}
\end{definition}

Given a client specification and a community of  available services, 
finding a simulation of the client by the community of  services 
amounts to constructing a winning strategy for \Eloise in the simulation game of the client 
by the asynchronous product of the available services. 
In the case of the PROM  example, a winning strategy for \Eloise can be computed 
in the  game $\mycal{G}(\ttt{CLIENT}, \ttt{FILE} \otimes \ttt{SEARCH})$,
and thus the client requests can be satisfied in all cases.

Note that the asynchronous
product of NFMAs cannot be defined easily  due to the global constraint
forcing the  registers values to be distinct ones (as discussed in  Subsec. \ref{subsec:gva:nfma:equiv}).
On the contrary, this construction is easy to specify with GVAs.


%% file: example.tex
\begin{figure}[h*]
\centering
  \vspace{-17pt}
  \begin{center}
    \scalebox{0.8}{\begin{tikzpicture}[shorten >=1pt,node distance=1.88cm, bend angle=60,
          on grid,auto, initial text=, >=stealth] 
        \node[state,initial] (p_0)   {$p_0$}; 
        \node[state] (p_1) [below =of p_0] {$p_1$}; 
        \node[state](p_2) [below =of p_1] {$p_2$};
        \node[state](p_3) [below  =of p_2] {$p_3$};
        \node[state](p_4) [below  =of p_3] {$p_4$};
        \node[state](p_5) [left  =of p_1] {$p_5$};

        \path[->] 
        (p_0) edge  node [left] { \ttt{Open(x)} } (p_1)
        (p_1) edge   node  [left] {\ttt{Open(y)}} (p_2)

        (p_2) edge  [bend right] node [right] {\ttt{Fail}}  (p_1)
        (p_1) edge [bend right]  node [right] { \ttt{Fail}} (p_0)

        (p_2) edge  node [right] { \ttt{Search(z)}} (p_3)

        (p_3) edge  [bend angle =40,bend left] node [left] {\ttt{Fail}} (p_2)

        (p_3) edge   node [left] {\ttt{Type(z,w)}} (p_4)

        (p_4) edge [bend angle =100, bend right]  node [right] {\ttt{Write(z,y)}} (p_2)

        (p_2) edge  [out=70,bend angle =40, bend left] node [left] {\ttt{Close(x)}} (p_5)
        (p_5) edge  [bend angle =40, bend left] node [left] {\ttt{Close(y)}} (p_0)

         -- (0,0.8) node {CLIENT}
         -- (-0.6,-3.15) node {\ttt{x}$\neq$ \ttt{y}}
         -- (2.4,-6)  node { \ttt{w $\neq$ prom}}

         -- (-2.7,-5.6)  node { \ttt{Write(z,x)}}
         -- (-2.6,-5.99) node { \ttt{w = prom}};

        \draw[->] (-0.40,-7.5) .. controls +(left:1.9cm) and +(left:2cm) .. (-0.35,-3.88);    

        \begin{scope}[xshift=6.5cm, yshift=-1cm,node distance=2cm]
          \node[state] (q_0)   {$q_0$}; 

          \path[->] 
          (q_0) edge  [loop left] node [left] {\ttt{Open(m)}} ()
          (q_0) edge  [loop right] node [right] {\ttt{Close(m)}} ()
          (q_0) edge  [loop below] node [below] {\ttt{Write(m,n)}} ()
          (q_0) edge  [loop above] node [above] {\ttt{Fail}} ()
          -- (0,1.6) node {FILE};
        \end{scope}

        \begin{scope}[xshift=6.6cm, yshift=-4.6cm,node distance=2cm]
          \node[state,initial] (r_0)   {$r_0$}; 
          \node[state] (r_1) [below =of r_0] {$r_1$}; 
          \path[->] 
          (r_0) edge  [bend left]  node [right] {\ttt{Search(i)}} (r_1)
          (r_1) edge  [bend left,out=80,in=110]  node  [left] {\ttt{Type(i,j)}} (r_0)
          (r_1) edge   node [left] {{}} (r_0)
          -- (0,0.8) node {SEARCH}
          -- (-0.46,-1.06) node {\ttt{Fail}};
        \end{scope}
    \end{tikzpicture}}
  \end{center}
 \vspace{-20pt}
  \caption{PROM  example.}
  \vspace{-10pt}
 \label{main:example:fig}
\end{figure}
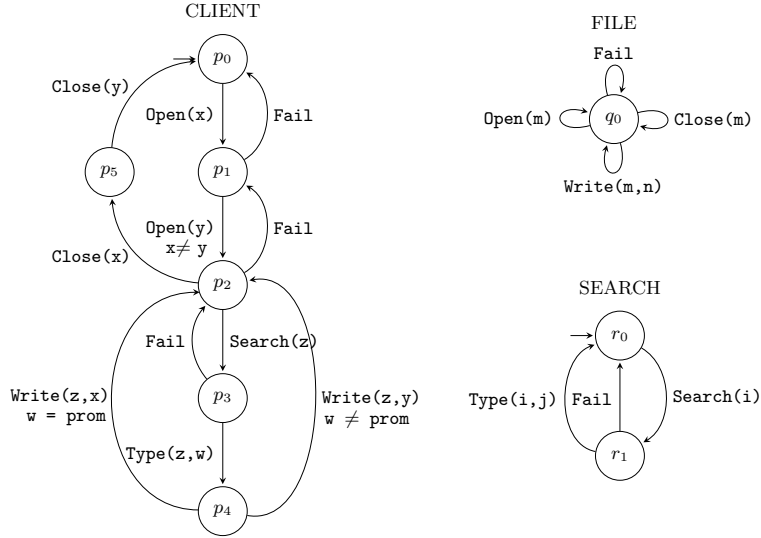

We illustrate the practical use of GVAs  through   
a service composition problem.
In Fig \ref{main:example:fig} we have an e-commerce  Web site allowing clients 
to open  files, search for  items  in a large domain 
that can be abstracted as infinite 
and  save them  to   an appropriate   file depending on the type  of the items 
(whether they are in promotion or not).
The three agents:  CLIENT,  FILE and SEARCH  communicate with 
messages ranging over a possibly  infinite set of terms. 
The problem is to check whether FILE and SEARCH can be composed in order to satisfy the 
CLIENT requests. Following ~\cite{BerardiCGP08,MW08} the problem reduces to find a simulation between 
CLIENT and the asynchronous product of FILE and SEARCH. 
The   variables $x$ and $y$ are  refreshed (i.e. freed to get a new value)
when  passing through the state $p_0$. In the same way  variables
$z$ and $w$ are   refreshed at  $p_2$. 
The variables $m$ and $n$  are refreshed at $q_0$; 
the  variables  $i$ and $j$ are refreshed at $r_0$. 
For saving  space,  a transition labeled 
by a term, say \ttt{write(m,n)},   abbreviates
 successive transitions labeled by the root symbol 
and its arguments, here \ttt{write}, \ttt{m} and \ttt{n}, respectively. 


%% file: conclusion.tex
\section{Conclusion}
\label{conclusion:sec}
In future works we plan to  investigate the decidability status  of the containment 
problem left open for the subclass of GVAs without disequalities in the guards.  
Our result on  GVAs simulation applies to the synthesis of web service composition.
In this context, disequalities should be useful to  express security policy enforcement on services in the spirit of ~\cite{DFM:nominal:CIAA12,BDF:MC:Usage:13}.




%% file: annex.tex
\appendix 
\section*{Appendices}
\label{annex}


 \setcounter{theorem}{0}
 \setcounter{proposition}{0}

\section{Proofs for Section \ref{gaut:sec}}
\label{req:sec:annex}

\LongVersion{
\begin{remark}
  \label{or:guards:rq}
  It is possible to consider the disjunction operation  ($\lor$) in the guards but one 
  can show that any GVA, in which the guards contain  the disjunction,  can be  turned 
  into a  GVA recognizing the same language and  
  in which the guards are without disjunctions, i.e. in the set $\G$ defined earlier.   
  But we must be careful about the instantiation of  the potential free variables  present in the guard:
  a transition $q_1\UberTo{\alpha,g_1\lor g_2} q_2$ can be equivalently turned into two transitions 
  $q_1\UberTo{\alpha,g_1 \land E_1} q_2$ and $q_1\UberTo{\alpha, g_2 \land E_2} q_2$,
  where $E_1= \bigland_{x \in S_1} (x=x)$
  (resp. $E_2= \bigland_{y \in S_2} (y =y)$) and $S_1  = \mycal{V}(g_2) \setminus \mycal{V}(g_1)$ 
  (resp. $S_2  = \mycal{V}(g_1) \setminus \mycal{V}(g_2)$).
\end{remark}
}

For two finite sets $A$ and $B$ we denote by $A^B$ the set of all 
total functions from $A$ to $B$.
Let $\mycal{X}$ and $\mycal{X}'$ be two disjoint sets of variables,
and let $\psi$ be a total function in $\mycal{X}^{\mycal{X}'}$, and let $g$ 
be a conjunction of equalities between variables in $\mycal{X}$.
Then define $g \sqsubset \psi$ iff there exists $x' \in \mycal{X}'$ s.t.
$\psi(x)=x'$ for all $x$ in $\mycal{V}(g)$.

\begin{lemma}(i.e. \emph{\textbf{Lemma}} \ref{GVA:GVAS:lemma})
For every GVA $\mycal{A}$ with $k$ variables and $n$ states 
there is a \GVAS  with $k+m$ variables and $O(n \cdot (k+m)!)$ states 
recognizing the same languages, where $m=|\Sigma_{\mycal{A}}|$.
\end{lemma}
\begin{proof}
Let $\mycal{A}=\model{\Sigma,\mycal{X},Q,Q_0,\delta,F,\kappa}$ be a GVA
with $\mycal{X}=\set{x_1,\ldots,x_k}$.

Firstly, we transform the GVA $\mycal{A}$ into a GVA $\bs{\mycal{A}}$
recognizing the same language and 
in which each  state is labeled with the set of variables   being free in this state. 
  We define 
  $\bs{\mycal{A}}=\model{\Sigma,\mycal{X},\bs{Q},\bs{Q_0},\bs{F},\bs{\delta},\bs{\kappa}}$ by:
  \begin{align*}
 \begin{cases}
  \bs{Q}  &= \set{ (q,X) \,\vert\, q\in Q \text{ and }X\subseteq      \mycal{X}}, \\
  \bs{Q_0} & = \set{ (q,\mycal{X}) \,\vert\, q\in Q_0 }, \\
  \bs{F} &= \set{ (q,X) \,\vert\, q\in F \text{ and }X\subseteq \mycal{X} }.
 \end{cases}
  \end{align*}
  The transition function $\bs{\delta}$ is defined by 
  $(q',X')\in \bs{\delta}((q,X),\alpha,g)$, where $\alpha \in \Sigma \cup \mycal{X}$ and $g$ is a guard,
  if  and only if, $q'\in \delta(q,\alpha,g)$ and  $X'=(X\setminus(\set{\alpha}\cup \mycal{V}(g)))\cup \kappa^{-1}(q')$.
  Finally, the refreshing function $\kappa'$ is defined by  $\bs{\kappa}(x)=
  \set{(q,X) \,\vert\, q\in\kappa(x)}$.

Secondly, we can assume w.l.o. 
that $\bs{\mycal{A}}$  has no constants and the variables are refreshed only in the 
states preceded  by  $\eps$-transitions. 
The constants  can be replaced by additional variables that
have  to be initialized with the related constants using an $\eps$-transition 
outgoing from the initial state. 
And, if some variables, say $X\subseteq \mycal{X}$, are refreshed in a state, say $\bs{q}$, 
then we add an $\eps$-transition $\bs{q}\uberTo{\eps} \bs{\tilde{q}}$ where 
the variables $X$ are refreshed in $\bs{\tilde{q}}$ instead of $\bs{q}$ and 
the outgoing transitions of $\bs{q}$ become the outgoing transitions of $\bs{\tilde{q}}$. 
 Thus, the guards of $\bs{\mycal{A}}$ are of the form
$\phi \land \phi'$ where $\phi$ (resp. $\phi'$) is a conjunction  of
equalities (resp. inequalities) between \emph{variables}. 
 
Thirdly, we let  $\bs{\mycal{A}}'$ to be the \GVAS 
$\bs{\mycal{A}}'=\model{\Sigma,\mycal{X}',\bs{Q}',\bs{Q}'_0, \bs{\delta}',\bs{F}',\bs{\kappa}'}$
defined by 
\begin{align*}
\begin{cases}
  \mycal{X}'  & =\set{x'_1,\ldots,x'_k} \\
  \bs{Q}'          &= \bs{Q} \times \mycal{X}^{\mycal{X}'} \\
  \bs{Q}'_0        & = \bs{Q}_0 \times \mycal{X}^{\mycal{X}'}  \\
  \bs{\kappa}'     & =  \kappa \times \mycal{X}^{\mycal{X}'}
\end{cases}
\end{align*}
and $\bs{\delta}'$ is defined by 
\begin{align*}
((q_1,X_1,\psi_1), (\alpha,\psi_1(g\land g')), (q_2,X_2,\psi_1)) \in  \bs{\delta}' \text{ iff } 
\begin{cases}
((q_1,X_1),(\alpha,g \land g'),(q_2,X_2)) \in \bs{\delta} \tand  \\
 \alpha \neq \eps \tand \\
 g   \sqsubset \psi_1  \tand \\
  \mycal{V}(g')  = codom(\psi_1) \tand \\
  \mycal{V}(g\land g') \cap X_1 =\emptyset
\end{cases}
\end{align*}
where $g$ (resp. $g'$) is a conjunction of equalities (resp. inequalities).
And, 
\begin{align*}
((q_1,X_1,\psi_1), (\eps, \psi_1(g \land g')), (q_2,X_2,\psi_2)) \in  \bs{\delta}' \text{ iff } 
\begin{cases}
((q_1,X_1),(\eps, g \land g'),(q_2,X_2)) \in \bs{\delta} \tand  \\
  \mycal{V}(g')  = codom(\psi_1) \tand \\
  \mycal{V}(g\land g') \subseteq  X_1 \tand \\
  \psi_2 =  \psi_1 \cup \set{x \mapsto x_0 \gvert x \in \mycal{V}(g)} \cup  \\
            \quad \quad  \quad  \set{x \mapsto y_0  \gvert x \in \mycal{V}(g')}  \\
       x_0= get (X' \setminus codom(\psi_1)) \\
       y_0= get \big(X' \setminus (codom(\psi_1) \cup \set{x_0})\big)
\end{cases}
\end{align*}
where $g$ (resp. $g'$) is a conjunction of equalities (resp. inequalities).
\qed
\end{proof}

\section{Proofs for Section \ref{prop:sec}}

The claims in the following Lemma are not hard to prove. They will be used in the proofs of 
the main claims. 
\begin{lemma}
  \label{synch:prop:lemma}
  Let $C\subseteq \Sigma$ be a finite set of  letters,  $\bar{\sigma}$ and $\sigma$  two substitutions, 
  $x$,    and $a$ a letter in $C$.
  The following hold.  
  \begin{enumerate}
  \item If  $\bar{\sigma} \synch_{C} \sigma$ then $|codom(\bar{\sigma})|= |codom(\sigma)|$. 
  \item If  $\bar{\sigma} \synch_{C} \sigma$  and   $D\subseteq Dom(\sigma)$ 
    then  $\bar{\sigma}_{|D} \synch_{C} \sigma_{|D}$, 
  \item  If $(\bar{\sigma}_1\uplus\bar{\sigma}_2) \synch  ({\sigma}_1\uplus {\sigma}_2)$ with 
    $dom(\bar{\sigma}_i)=dom(\sigma_i)$, then $\bar{\sigma}_i \synch \sigma_i$, for $i=1,2$. 
  \item If $\bar{\sigma} \synch_{C} \sigma$ and $\gamma$ is a substitution with $dom(\gamma)\cap dom(\sigma)=\emptyset$ and 
    $codom(\gamma)\subseteq C$, then 
    $\bar{\sigma}  \uplus \gamma \,\synch_{C}\, \sigma \uplus \gamma$.
  \item If $\bar{\sigma} \synch_{C} \sigma$ with $\bar{\sigma}(y)=\bar{a}$ and $\sigma(y)=a$ for some variable $y$, 
    and  $x\notin dom(\sigma)$ 
    then $\bar{\sigma}\uplus \set{x\mapsto \bar{a}} \synch_C \sigma \uplus \set{x\mapsto a}$.
  \item If $\bar{\sigma} \synch_{C} \sigma$ and $\bar{a} \notin C\cup codom(\bar{\sigma})$ 
    and $a \notin C\cup codom(\sigma)$ and $x\notin dom(\sigma)$ then 
    $\bar{\sigma}\uplus \set{x\mapsto \bar{a}} \synch_C \sigma \uplus \set{x\mapsto a}$.
  \end{enumerate}
\end{lemma}

Notice that the opposite direction of the Item $3$ of Lemma \ref{synch:prop:lemma}
does not hold in general.

\begin{lemma}
  \label{synch:subst:guards:lemma}
  Let $\sigma$ and $\bar{\sigma}$ be two substitution, where $\sigma \synch_{C} \bar{\sigma}$, 
  and let $g$ be a guard such that $\Sigma_{g} \subseteq C$.
  Then,  $\sigma \models g $ iff 
  $\bar{\sigma}\models g$.
\end{lemma}
\begin{proof}
  By induction on the structure of $g$ in both directions.
  \begin{trivlist}
  \item[$\Rightarrow$)] If $g=(\alpha=x)$, where $\alpha \in \Sigma$ and $x \in \mycal{X}$, then 
    $\sigma(x) \in C$, hence $\bar{\sigma}(x)  \in C$. Therefore $\sigma(x)=\bar{\sigma}(x)=\alpha$.
    If $g=(x=y)$, where  $x,y \in \mycal{X}$, then from the definition of $\synch_C$ we have that 
    $\sigma(x)=\sigma(y)$ iff $\bar{\sigma}(x)=\bar{\sigma}(y)$. Thus the claim holds.
    The case when  $g=g_1\land g_2$  follows from a direct application 
    of the induction hypothesis. 
  \item[$\Leftarrow$)] This direction follows from the fact that 
    $\sigma \synch_{C} \bar{\sigma}$ iff $ \bar{\sigma} \synch_{C} \sigma$. 
  \end{trivlist}
  \qed
\end{proof}

 For a substitution $\sigma$ and a guard $g$, 
  we shall write $\sigma  \vdash g$ if there exists a substitution $\sigma'$ 
  such that $\sigma \models \sigma'(g)$. Hence,
\begin{corollary}
  \label{synch:subst:guards:cor}
  Let $\sigma$ and $\bar{\sigma}$ be two substitution, where $\sigma \synch_{C} \bar{\sigma}$, 
  and let $g$ be a guard such that $\Sigma_{g} \subseteq C$.
  Then,  $\sigma \vdash g $ iff 
  $\bar{\sigma}\vdash  g$.
\end{corollary}
\begin{proof}$\text{ }$
  \begin{trivlist}
  \item[$\Rightarrow$)] 
    We show that if there exists a substitution $\gamma$ such that $dom(\gamma)=\Eu{V}(g) \setminus dom(\sigma)$ and 
    $\sigma \models  \gamma(g)$, 
    then  $\bar{\sigma}  \models  \gamma(g)$. But this follows from Lemma \ref{synch:subst:guards:lemma}.
  \item[$\Leftarrow$)] This direction follows from the fact that $\synch_C$ is symmetric relation.  
  \end{trivlist}
  \qed
\end{proof}

\begin{corollary}
  \label{synch:subst:guards:cor:2}
  Let $\sigma, \bar{\sigma}, \gamma,\bar{\gamma}$ be  substitutions, 
  where $dom(\gamma) \cap dom(\sigma)=\emptyset$  and $dom(\bar{\gamma}) \cap dom(\bar{\sigma})=\emptyset$.
  Let $g$ be a guard such that $\Sigma_{g} \subseteq C$.
  If $\sigma \uplus \gamma \synch_{C} \bar{\sigma} \uplus \bar{\gamma}$
  then we have that   $\gamma \models \sigma(g) $ iff 
  $ \bar{\gamma} \models \bar{\sigma}(g)$.
\end{corollary}

In what follows we 
let $S_1,S_2$ be two (possibly infinite) sets of letters with 
$|S_1\setminus S_2|>|\mycal{X}|$ and $|S_2 \setminus S_1|>|\mycal{X}|$ and $S_1\cap S_1\neq \emptyset$.
Let $C=S_1\cap S_2$.

\ucomment{In order to  relate the instantiations of the variables in the game $\mycal{G}$ to
the instantiation of the variables in the game $\overline{\mycal{G}}$, we need 
to introduce the function $\bs{\Theta}^{S_1,S_2}_{C}: \xi_{\mycal{X},S_1}  \times \xi_{\mycal{X},S_1} \times \xi_{\mycal{X},S_2}$, 
that, given three  substitutions  $M,\gamma$ and $M'$ 
such that $dom(M)\cap dom(\gamma)=\emptyset$ and $M  \synch_{C} M'$, constructs a substitution 
$\gamma'=\bs{\Theta}_{C}^{S_1,S_2}(M,\gamma,M')$ such that  $M \uplus \gamma \synch_{C} M' \uplus \gamma'$.}

\begin{definition}
\label{Theta:Def}
  We define the functions
  \begin{align*}
    \Theta_{C}^{S_1,S_2}, \bs{\Theta}_{C}^{S_1,S_2}: \xi_{\mycal{X},S_1}  \times \xi_{\mycal{X},S_1} \times \xi_{\mycal{X},S_2}
    \uberTo{} \xi_{\mycal{X},S_2}
  \end{align*}
  as follows.
  Let $M_1,\gamma_1 \in \xi_{\mycal{X},S_1}$, $M_2 \in \xi_{\mycal{X},S_2}$.
  Then, $\Theta_{C}^{S_1,S_2}(M_1,\gamma_1,M_2)$ 
  is   defined only when  $|dom(\gamma_1)|=1$ and $dom(\gamma_1)\cap dom(M_1)=\emptyset$ and $M_1\synch_{C} M_2$ by:
  \begin{align*}
    \Theta_{C}^{S_1,S_2}(M_1,\gamma_1,M_2)=
    \begin{cases}
      \gamma_1 & \tif \gamma_1(x) \in C \\ 
      \set{x\mapsto M_2(y)}      
      & \tif  \gamma_1(x) \in codom(M_1) \setminus C \tand \\
      & \;\;\quad M_1(y)=\gamma_1(x), y \in \mycal{X}\\
      \set{x\mapsto get(S_2\setminus codom(M_2))}      &  \tif \gamma_1(x) \in S_1 \setminus (C\cup codom(M_1))
    \end{cases}
  \end{align*}
  where  $dom(\gamma_1)=\set{x}$.\\
  And  
  $\bs{\Theta}_{C}^{S_1,S_2}(M_1,\gamma_1,M_2)$ is   defined  only when $dom(\gamma_1)\cap dom(M_1)=\emptyset$ by:
  \begin{align*}
    \bs{\Theta}_{C}^{S_1,S_2}(M_1,\gamma_1,M_2)=
    \begin{cases}
      \Theta_{C}^{S_1,S_2}(M_1,\gamma_1,M_2) & \tif |\gamma_1|=1\\
      &\\
      \gamma'_2 \uplus \bs{\Theta}_{C}^{S_1,S_2}(M_1 \uplus  \gamma'_1,  \gamma''_1, M_2\uplus \gamma'_2) 
      & \tif |\gamma_1|\ge 2,  \gamma_1=\gamma'_1\uplus \gamma''_1 \tand |\gamma'_1|=1, \\
      &\text{ where } 
      \gamma'_2=\Theta_{C}^{S_1,S_2}(M_1, \gamma'_1, M_2) 
    \end{cases}
  \end{align*}
\end{definition}

\begin{lemma}
  \label{Theta:prod:lemma}
  Let $M_1,\gamma_1 \in \xi_{\mycal{X},S_1}$ and $M_2 \in \xi_{\mycal{X},S_2}$ be 
  substitutions with  $dom(M_1)\cap dom(\gamma_1)=\emptyset$
  and $M_1 \synch_{C} M_2$.  We have that
  \begin{align*}
    (M_1 \uplus \gamma_1) \; \synch_{C}  \;  (M_2 \uplus \bs{\Theta}_{C}^{S_1,S_2}(M_1,\gamma_1,M_2))  
  \end{align*}
\end{lemma}
\begin{proof}
  By induction on $|dom(\gamma_1)|=1$. \\
  \begin{trivlist} 
  \item[\emph{Base Case.}] If  $|dom(\gamma_1)|=1$ then assume  $dom(\gamma_1)=\set{x}$ and
    let  $\gamma_2=\bs{\Theta}_{C}^{S_1,S_2}(M_1,\gamma_1,M_2)$. We distinguish three  cases depending on $\gamma_1(x)$. 
    \begin{itemize}
    \item If $\gamma_1(x) \in C$ then  it follows from the definition of 
      $\Theta_C^{S_1,S_2}$ that   $\gamma_2 =\gamma_1$.  From the Item 4  of Lemma \ref{synch:prop:lemma}  we get 
      $M_1\uplus  \gamma_1 \synch_C M_2\uplus  \gamma_1$.
    \item If $\gamma_1(x) \in codom(M_1)\setminus C$  then  in this case we recall that
      $\gamma_2=\set{x \mapsto M_2(y)}$ where 
      $M_1(y)=\gamma_1(x)$ for some variable  $y \in \mycal{X}$. The claim that \\
      $(M_1\uplus \set{x\mapsto \gamma_1(x)})  \synch_C M_2 \uplus \set{x \mapsto M_2(y)}$  follows from 
      the  Item 5  of Lemma \ref{synch:prop:lemma}.
    \item Otherwise, i.e. if $\gamma_1(x) \in  S_1 \setminus (C \cup codom(M_1))$ then the claim 
      that  \\ $(M_1\uplus \set{x\mapsto \gamma_1(x)})  \synch_C (M_2 \uplus \set{x \mapsto get(S_2\setminus codom(M_2))})$
      follows from the Item 6  of Lemma \ref{synch:prop:lemma}.
    \end{itemize}
  \item[\emph{Induction Case.}] Assume $\gamma_1=\gamma'_1 \uplus \gamma''_1$ with $|\gamma'_1|=1$. Let 
    \begin{align*}
      \begin{cases}
        \gamma'_2=\Theta_{C}^{S_1,S_2}(M_1, \gamma'_1, M_2),  \tand \\
        \gamma''_2 = \bs{\Theta}_{C}^{S_1,S_2}(M_1 \uplus  \gamma'_1,  \gamma''_1, M_2\uplus \gamma'_2) \tand  \\
        \gamma_2=\bs{\Theta}_{C}^{S_1,S_2}(M_1,  \gamma_1, M_2)= \gamma'_2 \uplus \gamma''_2
      \end{cases}
    \end{align*}
    From the induction 
    hypothesis  it follows that 
    \begin{align*}
      \begin{cases}
        M_1\uplus \gamma'_1  & \synch_C M_2 \uplus \gamma'_2 \\
        (M_1\uplus \gamma'_1)\uplus \gamma''_1   &  \synch_C (M_2 \uplus \gamma'_2)  \uplus \gamma''_2
      \end{cases}
    \end{align*}

    Therefore 
    \begin{align*}
      M_1\uplus \gamma_1    &  \synch_C  M_2  \uplus  \gamma_2
    \end{align*}
  \end{trivlist} 
  \qed
\end{proof}

\begin{lemma}(i.e. \emph{\textbf{Lemma}} \ref{lemma:finite:constants})
  Let $\mathcal{A}$ be a GVA  over $\Sigma$ with $n$ variables and $k$ constants
  $\Sigma_{\mycal{A}}=\set{c_1,\ldots,c_k}$. Let  $\bs{\Sigma}=\set{a_1,\ldots,a_n,c_1,\ldots,c_k}$. 
  Then, $\mycal{A}$ recognizes a non-empty language  over
  $\Sigma^\star$ if, and only if, it recognizes a non-empty language
  over $\bs{\Sigma}^{\star}$.
\end{lemma}
\begin{proof}
  Let $C=\set{c_1,\ldots,c_n}$.
  We show that there is a run 
$(\sigma_0,q_0)\uberTo{} \ldots \uberTo{}(\sigma_n,q_n)$  over  $\Sigma^{\star}$ in $\mycal{A}$ 
  iff  there is a run 
 $(\sigma'_0,q_0)\uberTo{} \ldots \uberTo{}(\sigma'_n,q_n)$ over   
 $\bs{\Sigma}^{\star}$ in $\mycal{A}$
  such that $\sigma_i \synch_{C} \sigma'_i$ for all $i=0,\ldots,n$. 
The proof is by induction
on $n$ in both directions. The base case $n=0$ holds trivially since $\sigma_0=\sigma'_0=\emptyset$. Assume that 
the claim holds up to $n$ and let us prove it for $n+1$.

\begin{itemize}
\item[{$\Rightarrow$)}] 
  Assume there is a transition $q_n \uberTo{\alpha_n,g_n} q_{n+1}$ in $\mycal{A}$ where $\alpha_n \in \Sigma \cup \mycal{X}$ and 
  $g_n$ is a guard.  From the induction hypothesis we have that $\sigma_n \synch_{C} \sigma'_n$. It follows 
  from  Lemma \ref{synch:subst:guards:lemma} that $\sigma_{n}(g_n)$ holds iff $\sigma'_{n}(g_n)$ holds. 
  Thus, the transition in $\mycal{A}$ over $\bs{\Sigma}$ is possible.  We describe next this transition.  
  From Definition \ref{run:lang:def} of the run of  GVAs, there 
  exists a substitution 
  $\gamma_n: \mycal{V}(\sigma_n(\alpha_n)) \cup \mycal{V}(\sigma_n(g_n)) \uberTo{} \Sigma$ such that 
   $(\gamma_n \uplus \sigma_n)(g_n)$ holds. 
 Hence, we must find a substitution 
 $\gamma'_n: \mycal{V}(\sigma'_n(\alpha_n)) \cup \mycal{V}(\sigma'_n(g_n)) \uberTo{} \bs{\Sigma}$ such that 
   $(\gamma'_n \uplus \sigma'_n)(g_n)$ holds. 
 We define $\gamma'_n$ by $\gamma'_n= \bs{\Theta}_{C}^{\Sigma,\bs{\Sigma}}(\sigma_n,\gamma_n,\sigma'_n)$.
 From Lemma \ref{Theta:prod:lemma} we have that $(\sigma_n \uplus \gamma_n) \synch_{C} (\sigma'_n \uplus \gamma'_n)$.
 Hence from Lemma \ref{synch:subst:guards:lemma} it follows that  $(\gamma'_n \uplus \sigma'_n)(g_n)$ holds.
 It remains to show that $\sigma_{n+1} \synch_{C} \sigma'_{n+1}$.
 But $\sigma_{n+1}=(\sigma_n \uplus \gamma_n)_{|D}$ and  $\sigma'_{n+1}=(\sigma'_n \uplus \gamma'_n)_{|D}$,
 for some set $D \subseteq  \mycal{X}$. From Item 9 of Lemma \ref{synch:prop:lemma} it follows that 
$\sigma_{n+1} \synch_{C} \sigma'_{n+1}$. 

\item[$\Leftarrow$)] Same proof but we call the function $\bs{\Theta}_{C}^{\bs{\Sigma},\Sigma}(\sigma'_n,\gamma'_n,\sigma_n)$.
\end{itemize} 
\qed
\end{proof}

\LongVersion{
\begin{definition}
  Let $D \subseteq \mycal{X}$ and $S \subset \G$ be a finite set of atomic guards, 
  and let $G=\bigland S$.
  The \emph{deletion} of  the variables in $D$ from  $G$, denoted by $G^{D}$ is  defined by
  $G^{D}=\bigland \set{g \in S \gvert \mycal{V}(g) \cap D =\emptyset }$.
\end{definition}
}

\section{Proofs for Section \ref{prop:sec}}
 \label{prop:sec:annex}

\subsection{Closure properties  of  GVAs}
\label{closure:sec:annex}

\setcounter{theorem}{0}
\begin{theorem}
\label{closure:prop:annex}
GVAs are closed under  concatenation, Kleene operator and
intersection. 
\end{theorem}
\begin{proof}
 Let  $\mathcal{A}_1=\model{\Sigma_1,\mycal{X}_1,Q_1,q_0^1,\delta_1,F_1,\kappa_1}$ 
and $\mathcal{A}_2=\model{\Sigma_2,\mycal{X}_2,Q_2,q_0^2,\delta_2,  F_2,\kappa_2}$ be two  GVAs.
Up to variable renaming  it is sufficient to consider the closure under the above operations
for two  GVAs that do not share variables.

 The closure under  Kleene operation and
concatenation  holds since    GVAs have  
 $\varepsilon$-transitions.
More precisely, the Kleene closure $\mycal{A}_1^{\star}$ amounts to adding an (unguarded) $\eps$-transition 
between the accepting states and initial states of $\mycal{A}_1$. And the concatenation
$\mycal{A}_1\cdot \mycal{A}_2$ amounts to adding  an (unguarded) $\eps$-transition between 
the accepting states of $\mycal{A}_1$ and the initial states of $\mycal{A}_2$.

The closure under intersection follows from the fact  that 
the intersection of two GVAs $\mycal{A}_1$  and $\mycal{A}_2$ 
denoted by $\mycal{A}_1\cap  \mycal{A}_2$  can be  defined as follows:
\begin{align*}
  \mycal{A}_1 \cap  \mycal{A}_2=\model{ \Sigma_1 \cup  \Sigma_2, \mycal{X}_1 \cup \mycal{X}_2, Q_1\times Q_2, q_0^1 \times q_0^2,\delta, F_1 \times F_2,\kappa},
\end{align*}
where $\delta$ and $\kappa$ are defined by:
 \[
\begin{cases}
(q'_1,q'_2) \in \delta((q_1,q_2),(\alpha_1, (\alpha_1=\alpha_2) \land g_1 \land g_2)) \quad \textrm{iff} \quad & q'_1 \in \delta_1(q_1,\alpha_1,g_1)
     \textrm{ and }  \\
   &  q'_2 \in \delta_2(q_2,\alpha_2,g_2).  \\
     (q_1,q_2) \in \kappa(x)  \quad  \textrm{iff} \quad   q_1 \in \kappa_1(x) \tor  q_2 \in \kappa_2(x).
\end{cases}
 \]
The proof that $L(\mycal{A}_1) \cap L(\mycal{A}_2) = L(\mycal{A}_1\cap \mycal{A}_2)$ is straightforward.
\qed
\end{proof}

\begin{proposition}
\label{complement:prop}
  The complement of a regular language is GVA-recognizable.  
  That is, given a FA $F$ there exists a GVA $\mycal{A}$ such that
  $L(\mycal{A})= \Sigma^{\star} \setminus L(F)$.
\end{proposition}
\begin{proof}
  The construction of $\mycal{A}$ is similar  to the one for FAs
  (over a finite alphabet). We assume that  $F$ is deterministic. 
  Firstly, we make the completion of $F$, i.e. we construct 
  an equivalent GVA so that 
  for each state $q$ of $F$ and for each letter $l \in \Sigma$ there is a unique  transition
  outgoing from $q$ that reads $l$. Secondly, we swap the accepting and non-accepting states.   

  Formally, assume $F=\model{\Sigma,Q,p_0,\delta,F}$, with $Q=\set{q_1,\ldots,q_n}$,
  and define
  $\mycal{A}=\model{\Sigma,\mathcal{X},Q',p_0,\delta',F',\kappa}$ by
  $$
  \left\lbrace
  \begin{array}{rl}
    \mathcal{X} &=\set{x_1,\ldots,x_n}\\
    Q' &= Q\cup \set{q_{i1},q_{i2}, i=1,\ldots,n}\\
    F' &= (Q \setminus F) \cup \set{q_{i1},q_{i2}, i=1,\ldots,n} \\
    \delta' &= \delta \cup \big\{q_i \UberTo{x_i, \bigland_j(x_i\neq a_j)} q_{i1} \gvert \text{ for all } a_j \tst q_i \uberTo{a_j} q_{i'} \in \delta  \big\} \cup  
    \big\{ q_{i2} \uberTo{x_i} q_{i2} \big\}\\
    \kappa(x_i)&=\set{q_{i2}}
  \end{array}
  \right.
  $$
  Notice that  $F$ rejects  a word $w$ iff $\mycal{A}$ accepts  $w$.
\end{proof}

\subsection{Decision procedures for GVAs}

\begin{theorem}
  \label{univ:GVA:th}
  For GVAs, Membership is NP-com\-plete. 
\end{theorem}
\begin{proof}
  Let $\mycal{A}$ be a GVA and $w=w_1 \cdots w_n$ a word in  $\Sigma^{\star}$. 

  For the upper  bound of the membership, a non deterministic polynomial algorithm 
  guesses a   path  in  $\mycal{A}$ of length $|w|$ such that the final state is accepting; 
  and a series of substitutions   $\sigma_1,\ldots, \sigma_{|w|}$, where $\sigma_i: \mycal{X}\uberTo{} \set{w_j,\, 1 \le j \le  |w|}$, 
  then checks wether  the corresponding  run on $w$ is possible.
  The lower bound, i.e. the NP-hardness, follows from the fact that the 
  membership problem for GVAs without guards, i.e. GVAs, is NP-complete\cite[Theorem 3]{BCR:FVA:13}. 
  The undecidability of the universality  follows from \cite{GKS10} since 
  this problem is undecidable for the class of variable automata which is a subclass of GVAs. 
  \qed 
\end{proof}

\section{Proofs for Section \ref{simulation:sec}}
\label{sim:sec:annex}

\ucomment{
The definition of simulation preorder for GVAs follows.
\begin{definition}
  \label{sim:GVA:def}
  Let   $\mycal{A}_1=\model{\Sigma,\mathcal{X}_1,Q_1,q^{1}_0,\tau_1,\delta_1,F_1,\kappa_1}$
  and  $\mycal{A}_2=\model{\Sigma,\mathcal{X}_2, \\ Q_2,q^2_0,\tau_2,\delta_2,F_2,\kappa_2}$ be two GVAs where
  $\mycal{X}_1 \cap \mycal{X}_2=\emptyset$.
  A simulation of  $\mycal{A}_1$ by $\mycal{A}_2$ is a relation 
  $\unlhd  \subseteq (\zeta_{\mycal{X}_1,\Sigma} \times Q_1) \times (\zeta_{\mycal{X}_2,\Sigma} \times Q_2)$ such that:
  \begin{itemize}
  \item $(\tau_1,q^1_0) \unlhd (\tau_2,q^2_0)$.
  \item if $(\sigma_1,q_1) \unlhd (\sigma_2,q_2)$ and if $(\sigma_1,q_1) \uberTo{a} (\sigma'_1,q'_1)$ for $a\in \Sigma$ 
    then there exists a state $q'_2 \in Q_2$ such that   $(\sigma_2,q_2) \uberTo{a} (\sigma'_2,q'_2)$ and 
    $(\sigma'_1,q'_1) \unlhd (\sigma'_2,q'_2)$. 
  \end{itemize}
\end{definition}
}


The definition of the function $\Xi$ follows.

\begin{lemma}
  \label{xi:lemma}
  Let  $\sigma_1,\sigma_2,\sigma_3,\gamma_2 \in \xi_{\mycal{X},S_1}$ and $\alpha,\beta \in C \cup \mycal{X}$ 
  and  $\sigma'_1,\sigma'_2,\sigma'_3 \in \xi_{\mycal{X},S_2}$  and $g_2 \in G$ be such that 
  \begin{align}
    \label{Xi:require:eq}
    \left\lbrace
    \begin{array}{rl}
      \gamma_2(\sigma_2(\beta)) &= \sigma_3(\alpha)   \\
      \sigma_3  \uplus \sigma_2 & \synch_{C}  \sigma'_3 \uplus \sigma'_2 \\
      \sigma_1 & \synch_C \sigma'_1 \\  
      \gamma_2  &\models \sigma_2(g_2)  \\
      dom(\sigma_1) & \subseteq dom(\sigma_3)  \\
      dom(\sigma'_1)  & \subseteq dom(\sigma'_3) \\
    \end{array}
    \right.
  \end{align}  

  There exists a  function $\Xi_{C}^{S_1,S_2}$
  \begin{align*}
    \Xi_{C}^{S_1,S_2}: \quad  \xi_{\mycal{X},S_1}^{5} \times (C \cup \mycal{X})^2    \times G_{C}  \times \xi_{\mycal{X},S_2}^{3}  & \uberTo{} \xi_{\mycal{X},S_2}\\
    (\sigma_1,\sigma_2,\sigma_3,\gamma_2,\alpha,\beta,g_2,\sigma'_1,\sigma'_2,\sigma'_3) & \mapsto \gamma'_2
  \end{align*}

  which is defined only if Eq (\ref{Xi:require:eq}) holds and satisfies the following:
  \begin{align}
    \gamma'_2(\sigma'_2(\beta)) &= \sigma'_3(\alpha)   \label{goal:Xi:1}  \tag{A.1}\\
    (\sigma_1 \uplus \gamma_1)  \uplus (\sigma_2 \uplus \gamma_2)  & \synch_{C} (\sigma'_1 \uplus \gamma'_1)  \uplus (\sigma'_2 \uplus \gamma'_2) \label{goal:Xi:2}  \tag{A.2} \\
    \gamma'_2  & \models \sigma'_2(g_2) \label{goal:Xi:4}  \tag{A.3}
  \end{align}  
\end{lemma}
\begin{proof}
  The construction of $\Xi_{C}^{S_1,S_2}$ depends on $\sigma_3(\alpha)$
  \begin{enumerate}[{I.)}]
  \item If $\sigma_3(\alpha) \in C$, then in  this case  we have 
    $\sigma_3(\alpha) = \sigma'_3(\alpha) \in C$. Hence $\sigma'_2=\sigma_2$.  Thus (\ref{goal:Xi:1}) holds.
    Furthermore we let: 
    \begin{align*}
      \gamma'_2  & = \Theta_{C}^{S_1,S_2}(\sigma_3\uplus \sigma_2, \gamma_2, \sigma'_3\uplus \sigma'_2)
    \end{align*}

    From  Lemma \ref{Theta:prod:lemma} it follows that 
    \begin{align*}
      \sigma_3\uplus \sigma_2 \uplus  \gamma_1 \uplus \gamma_2 & \synch_C \sigma'_3\uplus \sigma'_2 \uplus   \gamma'_1  \uplus \gamma'_2
    \end{align*}
    From Eq (\ref{Xi:require:eq}) we have $dom(\sigma_1) \subseteq dom(\sigma_3)$ and $dom(\sigma'_1)\subseteq dom(\sigma'_3)$
    and $dom(\sigma_1)=dom(\sigma'_1)$, hence it follows from Item 2 of Lemma \ref{synch:prop:lemma} that 
    \begin{align*}
      \sigma_1 \uplus \sigma_2  \uplus \gamma_2 & \synch_C \sigma'_1\uplus \sigma'_2  \uplus \gamma'_2
    \end{align*}
    Therefore   (\ref{goal:Xi:2}) holds. Finally 
    (\ref{goal:Xi:4}) follows from Corollary \ref{synch:subst:guards:cor:2}.

  \item If $\sigma_3(\alpha) \in S_1 \setminus C$, then
    $\alpha$ must be a variable,  say  $y_1 \in \mycal{X}$.
    We distinguish two cases depending on $\sigma_2(\beta)$. 
    \begin{enumerate}[{i.)}]
    \item If $\sigma_2(\beta)$ is a letter then in this case   
      $\sigma_2(\beta)=\sigma_3(\alpha)$, and  we let 

      \begin{align*}
        \gamma'_2  & = \Theta_{C}^{S_1,S_2}(\sigma_3\uplus \sigma_2 \uplus \gamma_1, \gamma_2, \sigma'_3\uplus \sigma'_2 \uplus \gamma'_1)
      \end{align*}

      And we must show  that   $\sigma'_3(\alpha)=\sigma'_2(\beta)$. 
      Notice that $\beta$  must be a  variable, say $y_2\in \mycal{X}$. 
      On the one hand we have that  $\set{y_1 \mapsto \sigma_3(\alpha), y_2 \mapsto \sigma_2(\beta)} \subseteq \sigma_3\uplus \sigma_2$
      and   $\set{y_1  \mapsto \sigma'_3(\alpha), y_2  \mapsto \sigma'_2(\beta)} \subseteq \sigma'_3 \uplus \sigma'_2$. 
      On the other hand, we have that  $\sigma_3(\alpha)=\sigma_2(\beta)$  and $\sigma_3\uplus \sigma_2 \synch_C \sigma'_3\uplus \sigma'_2$.
      Therefore  $\sigma'_3(\alpha)= \sigma'_2(\beta)$, thus (\ref{goal:Xi:1}), (\ref{goal:Xi:2})  and (\ref{goal:Xi:4}) hold.

    \item If $\sigma_2(\beta)$ is a variable, say $y_2\in \mycal{X}$, then $\sigma'_2(\beta)=\sigma_2(\beta)=\beta=y_2$, since  
      $\sigma'_2 \synch \sigma_2$.     In this case   we have $\set{y_2 \mapsto \sigma_3(\alpha)} \subseteq \gamma_2$. 
      Thus we let
      \begin{align*}
        \gamma'_2 & = \Theta_{C}^{S_1,S_2}(\sigma_3\uplus \sigma_2  \uplus \set{y_1\mapsto \sigma_3(\alpha)}, 
        \gamma_2, \sigma'_3\uplus \sigma'_2 \uplus  \set{y_1\mapsto \sigma'_3(\alpha)})
      \end{align*}
      And Eqs (\ref{goal:Xi:1}), \ldots,  (\ref{goal:Xi:4}) hold.
    \end{enumerate}
    
  \item If $\sigma_3(\alpha)$ is a variable, say $x_1\in \mycal{X}$, then $\sigma'_3(\alpha)=\sigma_3(\alpha)=\alpha=x_1$. 
    We distinguish two cases  
    depending on the nature of $\sigma_2(\beta)$. 
    \begin{enumerate}[{I.)}]
    \item If $\sigma_2(\beta)$ is a letter  then $\sigma'_2(\beta)$ is a letter as well since 
      $\sigma'_2 \synch \sigma_2$. 
      This case is dual w.r.t. case II.ii). 
    \item  If $\sigma_2(\beta)$ is a variable, say $y_2\in \mycal{X}$, then  
      $\sigma_2(\beta)=\sigma'_2(\beta)=\beta=y_2$ since $\sigma'_2 \synch \sigma_2$.
      In this case we let $\gamma'_2=\Theta_{C}^{S_1,S_2}(\sigma_1\uplus \sigma_2, \gamma_2,\sigma'_1\uplus \sigma'_2)$.
    \end{enumerate}
  \end{enumerate}
  \qed
\end{proof}

By using the functions $\bs{\Theta}$ and $\Xi$ and their respective 
properties stated in Lemmas 
\ref{Theta:prod:lemma} and \ref{xi:lemma} 
we are ready to prove Lemma \ref{f:construct:simul:lemma}

\begin{lemma} (i.e. \emph{\textbf{Lemma}} \ref{f:construct:simul:lemma})
\label{f:construct:simul:lemma:annex}
Let $\mycal{A}_1=\model{\Sigma_1,\mathcal{X}_1,Q_1,Q_0^1,\delta_1,F_1,\kappa_1}$ and 
$\mycal{A}_2=\model{\Sigma_2,\mathcal{X}_2,Q_2,Q_0^2,\delta_2,F_2,\kappa_2}$ be two GVAs, and let 
 $\mycal{X}=\mycal{X}_1 \cup \mycal{X}_2$.
Let $\mycal{G}(\mycal{A}_1,\mycal{A}_2)$ (resp.  
 $\overline{\mycal{G}}(\mycal{A}_1,\mycal{A}_2)$)  be the  simulation game in which the two players 
 instantiate the variables from $\Sigma$ (resp. $C_0$ defined  in Eq. (\ref{C0:Def})).
 Let $p^\star$ and $\overline{p}^{\star}$ be their starting position respectively.
Then, there is a   function  $f$:
\begin{align*}
f: \, Pos(\mycal{G}(\mycal{A}_1,\mycal{A}_2)) \uberTo{} Pos(\overline{\mycal{G}}(\mycal{A}_1,\mycal{A}_2))
\end{align*}
 with $f(p^\star)=\overline{p}^{\star}$ and $\wp \synch_{C} f(\wp)$ for all $\wp \in Pos(\mycal{G}(\mycal{A}_1,\mycal{A}_2)$,
   such that    the  following hold:  
\begin{enumerate}[i)]
\item  for all $\overline{\wp} \in  \Pos_{A}(\overline{\mycal{G}}(\mycal{A}_1,\mycal{A}_2))$, 
       if $\overline{\wp} \uberTo{} \overline{\wp}'$ is a move of \Abelard in $\overline{\mycal{G}}(\mycal{A}_1,\mycal{A}_2)$ 
       and $f(\wp)=\overline{\wp}$ for some  position $\wp$ in $\mycal{G}(\mycal{A}_1,\mycal{A}_2)$, 
       then there exists a position $\wp'$ in $\mycal{G}(\mycal{A}_1,\mycal{A}_2)$  such that the move 
       $\wp \uberTo{} \wp'$ is possible in $\mycal{G}(\mycal{A}_1,\mycal{A}_2)$ and $f(\wp')=\overline{\wp}'$. And, 
\item for all $\wp \in  \Pos_{E}(\mycal{G}(\mycal{A}_1,\mycal{A}_2))$, 
       if $\wp \uberTo{} \wp'$ is a move of \Eloise  in $\mycal{G}(\mycal{A}_1,\mycal{A}_2)$ 
       then there exists a position $\overline{\wp}'$ in $\overline{\mycal{G}}(\mycal{A}_1,\mycal{A}_2)$  such that the move 
       $f(\wp) \uberTo{} \overline{\wp}'$ is possible in $\overline{\mycal{G}}(\mycal{A}_1,\mycal{A}_2)$ and $f(\wp')=\overline{\wp}'$.
\end{enumerate}
\end{lemma}
\begin{proof}
  The proof is by induction on $n$,  the number  of the  moves  
  made in $\overline{\mycal{G}}(\mycal{A}_1,\mycal{A}_2)$  plus the number of moves made in  ${\mycal{G}}(\mycal{A}_1,\mycal{A}_2)$.
  The base case, i.e. when $n=0$,   trivially holds  since the starting position of 
  $\overline{\mycal{G}}(\mycal{A}_1,\mycal{A}_2)$   and of $\mycal{G}(\mycal{A}_1,\mycal{A}_2)$ is 
  $\AbelardM{(\emptyset,q_0^1),(\emptyset,q_0^2)}$.
  
  For the induction case let $\wp_n \in \Pos(\mycal{G}(\mycal{A}_1,\mycal{A}_2))$  and $\bar{\wp}_n \in \Pos(\overline{\mycal{G}}(\mycal{A}_1,\mycal{A}_2))$ where 
   $f(\wp_n)=\bar{\wp}_n$.  
  We discuss  two cases depending whether   $\bar{\wp}_n$ and 
  $\wp_n$  are both \Abelard positions   or  they are both \Eloise positions. 

  \begin{itemize}
  \item[\emph{i)}]  If $\wp_{n} \in \PosA(\mycal{G}(\mycal{A}_1,\mycal{A}_2))$ and 
    $\bar{\wp}_{n} \in \PosA(\overline{\mycal{G}}(\mycal{A}_1,\mycal{A}_2))$ then 
       consider an  \Abelard move $\bar{m}=\bar{\wp}_n \uberTo{} \bar{\wp}_{n+1}$ in 
    $\overline{\mycal{G}}(\mycal{A}_1,\mycal{A}_2)$. 
    In this case  $\bar{m}$ is of the form:
    \begin{align*}
      \bar{m} &= 
      \underbrace{\AbelardM{(\bar{\sigma}_1,q_1), (\bar{\sigma}_2,q_2)}}_{\bar{\wp}_n}
      \uberTo{} 
            \underbrace{\EloiseM{({(\bar{\sigma}_1\uplus \bar{\gamma})}_{{|D}},q'_1),  (\bar{\sigma}_2,q_2)}{(\bar{\sigma}_1\uplus \bar{\gamma},g_1)}}_{\bar{\wp}_{n+1}}\\
      &\hspace*{1cm}\vert\hspace*{0.5cm} q'_1 \in \delta_1(q_1,\alpha,g_1)   \\
      &\hspace*{1.5cm}    \tand D=Dom(\bar{\sigma}_1\uplus \bar{\gamma}) \setminus \kappa_1^{-1}(q'_1) \\
      &\hspace*{1.5cm} \tand   \bar{\sigma}_1 \uplus \bar{\gamma} \vdash  g_1 \\ 
      &\hspace*{1.5cm} \tand    \bar{\gamma} : \mycal{V}(\bar{\sigma}_1(g_1)) \setminus \mycal{V}(\bar{\sigma}_1(\alpha)) \uberTo{} C_0
    \end{align*}

    From  the induction hypothesis  we have  
    $ \wp_n \synch_{C} f(\wp_n)$, that is,  $ \wp_n \synch_{C} \bar{\wp}_n $.  Hence   $\wp_n=\AbelardM{(\sigma_1,q_1),(\sigma_2,q_2)}$, for 
    two  substitutions $\sigma_1,\sigma_2$ 
    where   $(\bar{\sigma}_1 \uplus \bar{\sigma}_2) \synch_{C}(\sigma_1\uplus \sigma_2)$.
    Thus \Abelard   move $m$ in $\mycal{G}(\mycal{A}_1,\mycal{A}_2)$ is defined by:
    \begin{align*}
      m &= 
      \underbrace{\AbelardM{(\sigma_1,q_1), (\sigma_2,q_2)}}_{\wp_n}
          \uberTo{} 
      \underbrace{\EloiseM{({(\sigma_1\uplus \gamma)}_{{|D}},q'_1),  (\sigma_2,q_2)}{(\sigma_1\uplus \gamma)}}_{\wp_{n+1}}
    \end{align*}

    where $\gamma : \mycal{V}(\sigma_1(g_1)) \setminus \Eu{V}(\alpha) \uberTo{} \Sigma$ is defined by 
    \begin{align}
      \label{def:gamma:case:i:eq}
      \gamma=\bs{\Theta}^{C_0,\Sigma}_C( \bar{\sigma}_1\uplus \bar{\sigma}_2,  \bar{\gamma}, \sigma_1\uplus \sigma_2).
    \end{align}

    Notice that  since $\bar{\sigma}_1 \synch_{C}\sigma_1$ then $dom(\bar{\gamma})=dom(\gamma)$.
    We let $f(\wp_{n+1})=\bar{\wp}_{n+1}$
    Furthermore, we must show that 
    \begin{align}
      \label{goal:case:i:guard}
      \sigma_1 \uplus \gamma \vdash g_1
    \end{align}

    and  that $ \wp_{n+1} \synch_{C} f(\wp_{n+1}) $, i.e. to show that $\wp_{n+1} \synch_{C}  \bar{\wp}_{n+1}$. 
    That is, we must  show that: 
    \begin{align}
      \label{goal:case:i:eq}
      \big( {(\sigma_1\uplus \gamma)}_{{|D}} \cup (\sigma_1\uplus \gamma) \big) \uplus \sigma_2. 
      \; \synch_{C}\; 
      \big( {(\bar{\sigma}_1\uplus \bar{\gamma})}_{{|D}} \cup (\bar{\sigma}_1\uplus \bar{\gamma})  \big) \uplus \bar{\sigma}_2 
    \end{align}
    From the definition of $\gamma$ in  Eq (\ref{def:gamma:case:i:eq})  and by applying  Lemma \ref{Theta:prod:lemma} we get:
    $ (\sigma_1\uplus \sigma_2) \uplus \gamma     \; \synch_{C}\;    (\bar{\sigma}_1 \uplus \bar{\sigma}_2) \uplus \bar{\gamma}$.
     Therefore, 
    \begin{align}
      \label{intermezzo:relat:instantiation:eq}
      (\sigma_1 \uplus \gamma)  \uplus \sigma_2      \; \synch_{C}\;   
      (\bar{\sigma}_1 \uplus \bar{\gamma}) \uplus \bar{\sigma}_2 
    \end{align}
    On the one hand, it follows from the Item 3 of Lemma \ref{synch:prop:lemma} that  
    $(\sigma_1 \uplus \gamma)    \; \synch_{C}\;  (\bar{\sigma}_1 \uplus \bar{\gamma})$. Since we already have $\bar{\sigma}_1 \uplus \bar{\gamma} \vdash g_1$, 
    then it follows from Corollary \ref{synch:subst:guards:cor} that    $\sigma_1 \uplus \gamma \vdash g_1$.
    Thus Eq (\ref{goal:case:i:guard}) is proved. On the other hand, 
    since $M_{|D} \subseteq  M$ for any substitution $M$ and any $D\subseteq dom(M)$,  then   Eq (\ref{goal:case:i:eq}) 
    follows from Eq (\ref{intermezzo:relat:instantiation:eq}).

  \item[\emph{ii)}]   If $\wp_{n} \in \PosE(\mycal{G}(\mycal{A}_1,\mycal{A}_2))$ and 
    $\bar{\wp}_{n} \in \PosE(\overline{\mycal{G}}(\mycal{A}_1,\mycal{A}_2))$ then 
       consider  an  \Eloise move $m={\wp}_n \uberTo{} {\wp}_{n+1}$ in 
    ${\mycal{G}}(\mycal{A}_1,\mycal{A}_2)$. 

    In this case $m$ of the form: 
    \begin{align*}
      m&= \underbrace{\EloiseM{(\sigma_1,q_1),(\sigma_2,q_2)}{(\sigma_3,\alpha)}}_{\wp_n}
          \uberTo{} 
        \underbrace{\AbelardM{(\sigma_1 ,q_1), ((\sigma_2 \uplus \gamma_2)_{|D_2},q'_2)}}_{\wp_{n+1}}\\
      &\hspace*{1cm}\vert\hspace*{0.5cm}  q'_2 \in \delta_2(q_2,\beta,g_2)\\
      &\hspace*{1.5cm} \tand D_2=Dom(\sigma_2 \uplus \gamma_2) \setminus \kappa_2^{-1}(q'_2)\\
      &\hspace*{1.5cm} \tand \gamma_1(\sigma_3(\alpha))=\gamma_2(\sigma_2(\beta)) \\
      &\hspace*{1.5cm} \tand    \gamma_2 \models  \sigma_2(g_2) \\ 
      &\hspace*{1.5cm} \tand \gamma_2: \mycal{V}(\sigma_2(\beta)) \cup \mycal{V}(\sigma_2(g_2)) \to \Sigma
    \end{align*}

    From the induction hypothesis we have that $\wp_{n} \synch f(\wp_{n}) $, 
    that is,  $ \wp_{n} \synch_{C} \bar{\wp}_{n}$, therefore
    $\bar{\wp}_n= \EloiseM{(\bar{\sigma}_1,q_1), (\bar{\sigma}_2,q_2)}{(\bar{\sigma}_3,\alpha,g_1)}$, 
    for substitutions $\bar{\sigma}_1, \bar{\sigma}_2, \bar{\sigma}_3$,  
    such that $ (({\sigma}_1\cup {\sigma}_3) \uplus {\sigma}_2) \synch_{C} ((\bar{\sigma}_1 \cup \bar{\sigma}_3) \uplus \bar{\sigma}_2 )$.

    The corresponding move $\bar{m}$ in $\overline{\mycal{G}}(\mycal{A}_1,\mycal{A}_2)$
    is defined by: 
    \begin{align*}
      \bar{m}&= 
       \underbrace{\EloiseM{(\bar{\sigma}_1,q_1),(\bar{\sigma}_2,q_2)}{(\bar{\sigma}_3,\alpha)}}_{\bar{\wp}_n}
         \uberTo{} 
       \underbrace{\AbelardM{(\bar{\sigma}_1 ,q_1), ((\bar{\sigma}_2 \uplus \bar{\gamma}_2)_{|D_2},q'_2)}}_{\bar{\wp}_{n+1}}\\
      &\hspace*{1.5cm} \textrm{where } \bar{\gamma}_1(\bar{\sigma}_3(\alpha))=\bar{\gamma}_2(\bar{\sigma}_2(\beta)) \\
      &\hspace*{1.5cm} \tand    \bar{\gamma}_2 \models  \bar{\sigma}_2(g_2) \\ 
      &\hspace*{1.5cm} \tand \bar{\gamma}_2: \mycal{V}(\bar{\sigma}_2(\beta)) \cup \mycal{V}(\bar{\sigma}_2(g_2)) \to C_0
    \end{align*}

    where  $\bar{\gamma}_2$ is  defined by 
    \begin{align*}
      \bar{\gamma}_2= \Xi_{C}^{\Sigma,C_0}(\sigma_1,\sigma_2,\sigma_3,\gamma_2,\alpha,\beta,g_2,\bar{\sigma}_1,\bar{\sigma}_2,\bar{\sigma}_3)
    \end{align*}

    From  Eq (\ref{goal:Xi:2}) of    Lemma \ref{xi:lemma} we get 
    \begin{align*}
      \sigma_1 \uplus \sigma_2 \uplus \gamma_2   & \synch_{C}  
      \bar{\sigma}_1  \uplus \bar{\sigma}_2 \uplus \bar{\gamma}_2
    \end{align*}

    From the Item 2 of Lemma \ref{synch:prop:lemma} it follows that $\wp_{n+1}\synch_{C} f(\wp_{n+1})$,  i.e. 
     $\wp_{n+1}\synch_{C} \bar{\wp}_{n+1}$, since 
    \begin{align*}
      \sigma_1 \uplus (\sigma_2 \uplus \gamma_2)_{|D_2}   & \synch_{C}  
      \bar{\sigma}_1  \uplus (\bar{\sigma}_2 \uplus \bar{\gamma}_2)_{|D_2}
    \end{align*}
  \end{itemize}
  This ends the proof of the Lemma.
\qed
\end{proof}